\documentclass[aps,prx,twocolumn,showpacs,preprintnumbers,superscriptaddress]{revtex4}
\usepackage{amsmath}
\usepackage{amssymb}
\usepackage{graphicx}
\usepackage{dcolumn}
\usepackage{bm}
\usepackage{color}
\usepackage[colorlinks=true,linkcolor=blue,citecolor=blue,urlcolor=blue]{hyperref}
\usepackage{ulem}

\newtheorem{theorem}{Theorem}
\newtheorem{lemma}{Lemma}

\newcommand{\drop}[1]{}

\definecolor{BluBondi}{rgb}{0.00,0.58,0.71}
\definecolor{tangerine}{rgb}{0.944,0.522,0}
\definecolor{brown}{rgb}{0.633,0.156,0.156}

\newcommand{\editor}[2]{%
  \expandafter\newcommand\csname #1note\endcsname[1]{%
    \textcolor{#2}{(\textbf{#1:} {\it ##1})}}%
  \expandafter\newcommand\csname #1\endcsname[1]{%
    \textcolor{#2}{##1}}%
  \expandafter\newcommand\csname #1cancel\endcsname[1]{%
    \textcolor{#2}{\sout{##1}}}%
  \expandafter\newcommand\csname #1change\endcsname[2]{%
    \textcolor{#2}{\sout{##1} ##2}}%
  \newenvironment{#1text}{\color{#2}}{\color{black}}
}
\editor{AF}{BluBondi}
\editor{NM}{red}  
\editor{TC}{tangerine} 
\newcommand{\acronym}{Spectral-DFT}

\begin{document}
\title{Functional theory of the occupied spectral density}

\author{Andrea \surname{Ferretti}}
\email[corresponding author: ]{andrea.ferretti@nano.cnr.it}
\affiliation{Centro S3, CNR--Istituto Nanoscienze, 41125 Modena, Italy}
\author{Nicola \surname{Marzari}}
\affiliation{Theory and Simulations of Materials (THEOS)
             and National Centre for Computational Design and Discovery of Novel Materials (MARVEL),
           \'Ecole Polytechnique F\'ed\'erale de Lausanne, 1015 Lausanne, Switzerland}

%
%
\date{\today}

%
%
\begin{abstract}
We address the problem of interacting electrons 
in an external potential  
by introducing the occupied spectral density $\rho(\mathbf{r},\omega)$ as fundamental variable. 
First, we formulate the problem using an embedding framework, and prove a one-to-one correspondence between a $\rho(\mathbf{r},\omega)$ and the local 
dynamical external potential $v_{\text{ext}}(\mathbf{r},\omega)$ that embeds the interacting electrons into an open quantum system. Then, we use the Klein functional to 
($i$) define a universal functional of $\rho(\mathbf{r},\omega)$,
($ii$) introduce a variational principle for the total energy as a functional of $\rho(\mathbf{r},\omega)$, 
and ($iii$) formulate a non-interacting mapping of spectral self-consistent equations suitable for numerical applications.
At variance with time-dependent density-functional theory, this formulation 
aims at studying charged excitations and electronic spectra 
-- including electronic correlations --
with a functional theory; 
An explicit and formally correct description of all electronic levels could also lead to more  accurate approximations for the total energy.
\end{abstract}

%
%
\maketitle


\section{Introduction}
\label{sec:intro}

Density-functional theory (DFT)~\cite{Hohenberg-Kohn1964PR,Eschrig2003book}
has revolutionized the field of electronic-structure calculations both in its conceptual framework and in
its practical applications.
By recasting the Schr\"odinger equation for $N$ interacting electrons in an external (static) potential into 
a variational principle for the ground-state charge density, it has dramatically reduced the complexity of the 
electronic-structure problem,
at the expense of introducing  
a universal but unknown functional. 
The success of DFT is underscored by the vast output of research papers~\cite{Burke2015ARPChem} in many and diverse fields~\cite{vanNoorden2014Nature,Marzari2021NatMater}, going well beyond the original core community 
of condensed-matter practitioners and developers and with applications ranging from materials modelling to drug design. 
Empowered by DFT, efforts in high-throughput screening and computational discovery~\cite{Curtarolo2013NatMater,Marzari2016NatMater,Hautier2019CMS,Jain2016NatRevMater}, lately also combined with machine learning~\cite{Deringer2021ChemRev,Huang2023Science}, have become prevalent. 
Still, predictive accuracy is one of the challenges for the field~\cite{Marzari2016NatMater,Marzari2021NatMater}: while advances to the quality of the DFT exchange-correlation functionals are continuously brought forward~\cite{Truhlar2014PTRSA,Sun2015PhysRevLett,HeadGordon2016JCP,Medvedev2017Science,Kepp2017Science}, the target of 1 kcal/mol chemical accuracy 
remains elusive, especially for systems exhibiting significant electronic correlations and/or multi reference character, limiting the ability of becoming truly predictive for materials or molecular properties. 

The Kohn-Sham (KS)~\cite{Kohn-Sham1965PR} formulation of DFT partitions the unknown universal functional into the sum of the well-defined kinetic energy of the non-interacting KS electrons and the electrostatic Hartree contribution, pushing all the leftovers into an unknown exchange-correlation (xc) third term. Extensive theoretical work has elucidated some of the challenging features that the exact xc functional and potential have to display, including the existence of derivative discontinuities~\cite{Perdew-Levy1983PRL,Godby1987PRB,Perdew2017PNAS}, potential steps~\cite{Tempel2009JChemTheoryComput,Helbig2009JChemPhys,Hellgren2012PhysRevA,Ferretti2014PhysRevB}, and ultra non-local dependence~\cite{Ghosez1997PhysRevB,Vanderbilt1997PRL}. Importantly, even if the auxiliary KS non-interacting electrons are often taken as approximations to the real electronic states~\cite{Stefanucci-vanLeeuwen2013book,Martin-Reining-Ceperley2016book} (i.e., the Dyson orbitals, within the quasi-particle approximation, and with some valid arguments~\cite{Casida1995PRA,Chong2002JChemPhys} in support of this assumption), it is important to underscore~\cite{Gatti2007PhysRevLett,Martin-Reining-Ceperley2016book,Marzari2021NatMater} that there is no formal connection between the KS states and the electronic excitations (except for the highest occupied state in finite systems~\cite{Perdew-Levy1983PRL}), thereby making KS-DFT a theory of total energies, and not of electronic energies or of quasi-particles. 
Therefore, exact KS-DFT would reproduce the total energy (thermodynamics) of the electronic system without actually reproducing the individual electronic levels (charged excitations, or one-particle spectra).
Resorting to generalized KS schemes~\cite{Seidl1996PhysRevB,Garrick2020PhysRevX} (such as those involved in hybrid functionals~\cite{Perdew1996JCP,Heyd2003JCP,Brawand2016PRX}, meta-GGAs~\cite{Tao2003PRL,Sun2015PhysRevLett}, or DFT+U~\cite{Anisimov1991PhysRevB,Dudarev1998PhysRevB,Cococcioni-deGironcoli2005PRB,Timrov2018PRB}) may improve some predictions, such as those for band gaps and band widths~\cite{Fuchs2007PhysRevB,Marques2011PhysRevB,Kronik2012JChemTheoryComput,Ferretti2012PhysRevB,Brawand2016PRX,Perdew2017PNAS}, but
does not resolve this formal limitation (for instance, no satellites or quasi-particle renormalization factors can be obtained).

In order to determine simultaneously total energies and spectral properties, while remaining within the conceptual definition of DFT, is to introduce additional functionals that address specific observables, as recently proposed by the theory of connectors~\cite{Vanzini2020FaradayDiscuss,Vanzini2022npjComputMater}. 
Alternatively, one can move beyond standard DFT (e.g., with ensemble DFT~\cite{Gross-Oliveira-Kohn1988PRA,Gould2019PhysRevLett,Cernatic2021TopCurrChem,Gould2025PhysRevLett,Gould2025arXiv}, 
time-dependent DFT with unbound electrons~\cite{Giovannini2012PhysRevA}, or RDMFT~\cite{Sabatino2015JChemPhys}) to address charged excitations.
Koopmans' spectral functionals~\cite{Dabo2010PhysRevB,Linscott2023JChemTheoryComput} provide also a beyond-DFT functional approach to spectral properties~\cite{Ferretti2014PhysRevB}, very much in the spirit of the present paper.
On the other hand, Green's function (GF) methods~\cite{Stefanucci-vanLeeuwen2013book, Martin-Reining-Ceperley2016book} represent a formally straightforward and computationally viable alternative, since the one-particle GF naturally describes spectroscopic properties (charged excitations) and can be used to evaluate total energies either via the Galitskii-Migdal expression~\cite{Fetter-Walecka1971book,Martin-Reining-Ceperley2016book} or, variationally, using the Luttinger-Ward or Klein functionals~\cite{Luttinger-Ward1960PR,Klein1961PR,Baym1961PhysRev,Almbladh1999InternationalJournalofModernPhysicsB}.
Diagrammatic approximations to the self-energy include the GW method~\cite{Hedin1965PR,Reining2018wcms,Golze2019FIC}, possibly combined with different forms of vertex corrections~\cite{Ren2015PhysRevB,Chen-Pasquarello2015PRB,Kutepov2017PhysRevB,Maggio2017JChemTheoryComput,Vacondio2022JChemTheoryComput} and solved at different  levels of self-consistency~\cite{Reining2018wcms,Golze2019FIC}, providing the current state-of-the-art in terms of accuracy and cost. 
When projected on a localized manifold and combined with DFT, GW leads to the dynamical Hubbard approach~\cite{Chiarotti2023PhD,Chiarotti2024PRR} (featuring a dynamical $U$ term), showing remarkable results on both total energies and spectra, e.g., of transition-metal oxides~\cite{Chiarotti2024PRR,Caserta2025arXiv}.
Dynamical mean-field theory (DMFT)~\cite{Georges1996RevModPhys,Kotliar2006RevModPhys}, typically adopted to treat strongly correlated systems, is also able to address total energies and one-particle spectra and can be seen as a functional theory of the local Green's function~\cite{Savrasov2004PhysRevB} with approximations built from accurate solutions of the Anderson impurity model. 
While initially evaluated on top of DFT (e.g., at the LDA level)~\cite{Lichtenstein1998PhysRevB}, DMFT has also been combined with GW~\cite{Biermann2003PhysRevLett,Ayral2012PhysRevLett}.

In this work we take a fundamentally different route to treat excitations, spectra, and correlations, and
we propose a functional formulation of the electronic-structure problem
where the fundamental variable is the occupied spectral density $\rho(\mathbf{r},\omega)$~\cite{Martin-Reining-Ceperley2016book,Gatti2007PhysRevLett} 
(i.e., the diagonal of the hole spectral function $A(\omega)$), rather than the charge density $\rho(\mathbf{r})$. 
This approach avoids the complexity of dealing with functionals of the whole Green's function, 
and can be seen as a key step in a hierarchy of 
methods (see Box~1 of Ref.~[\onlinecite{Marzari2021NatMater}]) that move from DFT to GF-based approaches. While TDDFT~\cite{Runge-Gross1984PRL,Petersilka1996PRL} naturally addresses neutral excitations, the present approach  aims at charged excitations, in principle including all correlation features such as band renormalizations and satellites.

The reasons to adopt the occupied spectral density $\rho(\mathbf{r},\omega)$ as basic variable are manifold: 
($i$) The spectral density contains more information than the charge density, including one-particle energy levels (both quasi-particle peaks and satellites, and, in general, all non-coherent structures beyond them);
($ii$) in the KS spirit, this (interacting) spectral density can be obtained from a non-interacting system subject to a local and dynamical spectral potential~\cite{Gatti2007PhysRevLett}; 
This was hinted not to
need to exhibit a derivative discontinuity~\cite{Ferretti2014PhysRevB} 
and so avoids some of the exact 
constraints that the static KS potential needs to satisfy~\cite{Helbig2009JChemPhys,Tempel2009JChemTheoryComput};
($iii$) having an electronic-structure formulation where the single-particle energies correspond to real physical energies could lead naturally to more accurate total energies. 

Actually, more information needed to build the energy functional becomes available when using spectral densities, and more exact quantities become known. As an example (shown later), the interacting kinetic energy can be described exactly, provided some mild conditions are satisfied, once the spectral density is known. 
Notably, situations where a physical insulator is described by a metallic KS system~\cite{Godby1989PhysRevLett,Ghosez1997PhysRevB} are no longer expected to occur within this framework.

In order to build a functional theory of the (occupied) spectral density (\acronym), (1) we first extend the class of external potentials by developing a local embedding formulation of the electronic problem, 
in order to introduce effective dynamical and local external potentials 
originating from a bath coupled to the main system (some important technical results about GF embedding that are used within this work can be found in Ref.~\onlinecite{Ferretti2024PRB}).
Once this class of local and dynamical external potentials $v_{\text{ext}}(\mathbf{r},\omega)$ has been introduced, 
(2) we show that these potentials have a one-to-one mapping with the occupied spectral densities. This is the central result -- very much analogous to the first Hohenberg-Kohn theorem~\cite{Hohenberg-Kohn1964PR} -- that allows one to build a functional theory of the spectral density. Then, (3) the approach is equipped with a variational formulation; and last, (4) a non-interacting KS-like system, subject to a dynamical auxiliary potential, is introduced to represent the spectral density and provide a viable computational scheme to obtain it. 
So, with the present work we develop a functional theory of the spectral density by defining and proving a one-to-one correspondence and building a total energy functional.

In this context, an inspirational paper by Gatti et al.~\cite{Gatti2007PhysRevLett} proposed a spectral perspective, introducing a frequency-dependent Sham-Schl\"uter equation, later also cast -- in linearized form -- as a dynamical optimized effective potential (OEP) problem~\cite{Ferretti2014PhysRevB}. 
Also, the work by Savrasov and Kotliar~\cite{Savrasov2004PhysRevB}, later further developed~\cite{Biermann2003PhysRevLett,Leonov2014PhysRevLett,Haule2015PhysRevLett}, provides a variational functional of the local Green's function $G_{\text{loc}}$ (i.e. the diagonal of $G$ on some local orbital basis) underpinning DMFT~\cite{Georges1996RevModPhys,Kotliar2006RevModPhys}. The main difference with the present work is two-fold: first, we work strictly with the space-diagonal spectral function (i.e., the theory depends on a strictly local quantity). Most importantly, in this work only occupied states are concerned, making $\rho(\mathbf{r},\omega)$ a much simpler object than $G_{\text{loc}}(\omega)$.

In passing, we note that
the spectral density has also been discussed in a series of recent works developing the i-DFT framework~\cite{Jacob2018NanoLett,Jacob2020PhysRevLett,Sobrino2021PhysRevB}, a DFT extension where conceptual STS (scanning tunnelling spectroscopy) simulations can be run to extract the local spectral density.
Interesting, Aryasetiawan has recently proposed~\cite{Aryasetiawan2022PhysRevB,Aryasetiawan2025arXiv} a Green's function functional theory approach based on the introduction of a xc potential local in space and in time (related to a dynamical xc-hole) acting multiplicatively on the GF in lieu of the self-energy, opening the way to LDA-like approximations.
Finally, as argued in Ref.~[\onlinecite{Ferretti2014PhysRevB}], Koopmans functionals~\cite{Dabo2009arXiv,Dabo2010PhysRevB,Nguyen2018PhysRevX,Colonna2019JChemTheoryComput,Linscott2023JChemTheoryComput}, very successful in predicting spectral properties and leading to local and orbital-dependent potentials, can be seen as quasi-particle approximations to the \acronym{} functional.

The paper is organized as follows. In Sec.~\ref{sec:problem_def} we define the problem and introduce the concept of external potentials that are local and dynamical, together with some physical interpretation. Next, in Sec.~\ref{sec:one-to-one} 
we prove a one-to-one correspondence between occupied spectral densities and local, dynamical external potentials.
On the basis of this correspondence, in Sec.~\ref{sec:SF-universal} a universal functional of the spectral density is defined and a 
spectral density functional theory (\acronym) is formulated. Finally, in Sec.~\ref{sec:KS-mapping}, 
a non-interacting mapping, suitable for numerical applications, is introduced and discussed.


\section{Definition of the problem}
\label{sec:problem_def}
%
\subsection{A functional theory of the occupied spectral density}
\label{sec:spectral_density}
%
%
%
Given a system of interacting electrons, the spectral density $\rho(\mathbf{r},\omega)$ is the (spin integrated) diagonal of the imaginary part
of the one-particle Green's function (GF). By taking $\mathbf{x}\equiv\mathbf{r}\sigma$ as a space-spin coordinate, and using the Lehmann representation~\cite{Fetter-Walecka1971book,Stefanucci-vanLeeuwen2013book,Martin-Reining-Ceperley2016book}, it is
possible to write the GF as
\begin{equation}
   \label{eq:lehmann}
   G(\mathbf{x},\mathbf{x}',\omega) = \sum_s \frac{f_{s}(\mathbf{x}) \, f^{*}_s(\mathbf{x}')} 
                              { \omega -\epsilon_{s} +i\eta_s},
\end{equation}
where $f_s$ are the quasi-particle (QP) amplitudes or Dyson orbitals~\cite{Almbladh1985PhysRevB,Chong2002JChemPhys,Martin-Reining-Ceperley2016book,Golze2019FIC}:
\begin{eqnarray}
   \label{eq:amplitudes1}
   \epsilon_s < \mu & &\qquad f_s(\mathbf{x}) = \langle N-1,s | \, \hat{\psi}(\mathbf{x}) \, | N,0\rangle, \\
   \epsilon_s \ge \mu & &\qquad f_s(\mathbf{x}) = \langle N,0 | \, \hat{\psi}(\mathbf{x}) \, | N+1,s\rangle.
   \label{eq:amplitudes2}
\end{eqnarray}
Here $\epsilon_s=E^N_0-E^{N-1}_s$ ($\epsilon_s=E^{N+1}_s-E^{N}_0$) and $\eta_s=-i0^+$ ($\eta_s=+i0^{+}$)
for poles below and above the Fermi level $\mu$, respectively, 
where $| M,s\rangle$ and $E^M_s$ are the $s$-th excited state eigenvector and energy
of the $M$-particle system ($s=0$ for the ground state).

In this work we assume the poles of $G$ (and of any other time-ordered
frequency dependent operator, such as $\Sigma_{\text{Hxc}}$ or $v_{\text{ext}}$, see below) to be {\it discrete}. 
This can be formally achieved by considering the physical system of interest described by a finite box of arbitrary size
(e.g., with periodic-boundary conditions), meaning that a three-dimensional torus $\mathbb{T}^3$ 
is used as coordinate space instead of $\mathbb{R}^3$, as described in Ref.~[\onlinecite{Eschrig2003book}].
Despite some electronic-structure key quantities (e.g., bands or quasi-particle lifetimes) being rooted in the continuum of states, we do not foresee any major drawback in working with discrete quantities. In fact, computer simulations routinely solve the electronic-structure problems in the discrete (e.g., $\mathbf{k}$-point grids are used rather than continuous dispersions).
%
Moreover, we consider the Hamiltonian underlying the physical system to be not explicitly dependent on spin (i.e., no spin-orbit coupling nor external magnetic fields), so that the total spin $S_z$ is also a good quantum number. We also assume the ground state to be non-degenerate with $S_z=0$, so that $|N+1,s\rangle$ coupled to the ground state would only have $S_z=\pm 1/2$. Within these conditions we have a spin-collinear picture where $f_s(\mathbf{x})=f_{s}(\mathbf{r})$ with the spin index included in $s$, and
one also has $f_{s\uparrow}(\mathbf{r})=f_{s\downarrow}(\mathbf{r})$.

The occupied spectral density can then be written as:
\begin{eqnarray}
   \label{eq:spectral_density}
   \rho(\mathbf{r},\omega) &=& \sum_s^{\text{occ}} | f_s(\mathbf{r}) |^2 \, \delta(\omega -\epsilon_s)
   \\
   \label{eq:G_imag}
   &=& \frac{1}{\pi} \sum_\sigma \text{Im}[G_\sigma(\omega)]_{\mathbf{r},\mathbf{r}} \,\theta(\mu-\omega) \;,
   \\
   \label{eq:G_lesser}
   &=& \frac{1}{2 \pi i} \sum_\sigma \, G^{<}_\sigma(\mathbf{r},\mathbf{r},\omega)\, ,
 \end{eqnarray}
with $G^{<}(1,2) = i\langle \hat{\psi}^\dagger(2)\hat{\psi}(1)\rangle 
$
~\cite{Stefanucci-vanLeeuwen2013book,Martin-Reining-Ceperley2016book}; 
the occupied spectral density could actually be labeled  $\rho^{<}(\mathbf{r},\omega)$, but we do not use this superscript to simplify the notation.
In general, $f_s(\mathbf{r})$ are neither orthogonal among themselves nor normalized to 1,
and there are in principle infinitely many contributions
already for the occupied states. 
Importantly, in view of 
\begin{equation}
n(\mathbf{r})=\sum_s^{\text{occ}} \rho_s(\mathbf{r}) = \sum_s^{\text{occ}}  | f_s(\mathbf{r}) |^2 = \int_{-\infty}^{\mu} \, d\omega \, \rho(\mathbf{r},\omega)
\end{equation}
the densities $\{\rho_s\}$ can be seen as a frequency partition of the charge density $n(\mathbf{r})$.
Since we want to build a functional theory of the occupied spectral density, we aim at writing the total energy of the problem, in analogy with DFT~\cite{Hohenberg-Kohn1964PR,Levy1979PNAS}, as
\begin{equation}
  \label{eq:FT_partition}
  E^{{\text{SF}}}[\rho] = \int d\mathbf{r}d\omega \,v_{\text{ext}} \rho + F^{\text{SF}}[\rho],
\end{equation}
with $\rho(\mathbf{r},\omega)$ being the occupied spectral density instead of the charge density $n(\mathbf{r})$ and $F$ a universal functional of $\rho$.
The class of external potentials appropriate to this development is discussed in Sec.~\ref{sec:potential}.

Further physical insight about this formulation can be obtained by performing a Fourier transform of the occupied spectral density to the time domain. Following Eq.~\eqref{eq:G_lesser} and the definition of $G^<$, one has 
\begin{eqnarray}
  \rho(\mathbf{r},\tau) &=& \frac{1}{2\pi} \sum_\sigma \, \langle \, \hat{\psi}^\dagger(\mathbf{r}\sigma,t_0)
                                       \hat{\psi}(\mathbf{r}\sigma,\tau + t_0) \, \rangle,  
\end{eqnarray}
showing that the frequency-dependence of the spectral density is actually related to a time delay between the creation and annihilation of a particle in the system, which in turn connects to the dynamics of $N-1$ particles (charged excitations).
This is at variance with the framework of time-dependent DFT (TDDFT)~\cite{Runge-Gross1984PRL,Petersilka1996PRL,Ullrich2012book}, where the main variable, the time-dependent density $n(\mathbf{r},t)$, can be written as
\begin{equation}
  n(\mathbf{r},t) = \sum_\sigma \, \langle \, \hat{\psi}^\dagger(\mathbf{r}\sigma,t) \hat{\psi}(\mathbf{r}\sigma,t) \, \rangle,
\end{equation}
the two field operators being evaluated at the same time, with the average being on the initial state, and exposing the dynamics of neutral excitations.
In this respect, a functional theory of the occupied spectral density (that we labeled \acronym) and TDDFT both represent functional formulations able to describe excitations; These are complementary (being naturally oriented to different kinds of excitations and spectroscopies); Importantly, a special focus of \acronym{} will also be the description of total energy and thermochemistry of correlated electrons~\cite{Chiarotti2024PRR,Caserta2025arXiv}.
Recently, advances in the TDDFT formulation have been  made~\cite{Giovannini2012PhysRevA} to address photoemission spectroscopy~\cite{Damascelli2003RMP}, 
imposing a boundary condition on the TDDFT simulation with an electron far from the system, thereby achieving a charged-excitation dynamics. Conversely, \acronym{} could be applied, e.g., to an anion system to access neutral excitations.

As already anticipated, this work is related to the formulation proposed by Savrasov and Kotliar in Ref.~[\onlinecite{Savrasov2004PhysRevB,Kotliar2006RevModPhys}], which focuses on the short range part of the Green's function 
as the basic variable of the theory, and results in a framework underpinning dynamical mean-field theory. 
Besides the strictly local approach, the main difference with the present \acronym{} formulation is the use in Ref.~[\onlinecite{Savrasov2004PhysRevB}] of the local Green's function over the whole frequency domain (i.e., involving both occupied empty states); instead, this work accomplishes a one-to-one correspondence based on the occupied (and strictly local) spectral  density alone.

\subsection{External potential}
\label{sec:potential}

Following Eq.~\eqref{eq:FT_partition},
the external potential that couples to the spectral density (a partition of the charge density) can be chosen to be more general than the static local potential of DFT. 
Indeed, while in DFT 
a one-to-one density-potential correspondence is obtained~\cite{Hohenberg-Kohn1964PR,Levy1979PNAS,Eschrig2003book} in the form $n(\mathbf{r})  \longleftrightarrow v_{\text{ext}}(\mathbf{r})$, 
in this work we aim at generalizing such correspondence to dynamical quantities as 
$\rho(\mathbf{r},\omega)  \longleftrightarrow v_{\text{ext}}(\mathbf{r},\omega)$. 
The reasons to make this extension are manifold:  working with local and dynamical potential is useful to ($i$) identify and prove a one-to-one correspondence between potentials and densities, and ($ii$) widen the domain of spectral densities for which the functional is defined; Notably, ($iii$) the use of dynamical external potentials is crucial to make a Kohn-Sham-like construction, i.e., to build a non-interacting system with the same spectral density of the interacting one. Indeed, as shown in Ref.~[\onlinecite{Gatti2007PhysRevLett}], a KS-like system for the spectral density requires a local and dynamical potential (such construction is simply not possible using a static and local potential). ($iv$) Generalizing the class of external potentials allows one to access a larger set of reference systems, possibly including cases simple enough to be solved numerically, that could be later used to build functionals of the spectral density.

As detailed in Sec.~\ref{sec:ham_embedding}, local dynamical potentials $v_{\text{ext}}(\mathbf{r},\omega)$ can be seen as arising from a local (i.e. point-by-point) embedding of the system of interest, which then formally becomes an open quantum system.
The auxiliary sub-system defining the quantum bath is assumed to be non-interacting, making the electron-electron interaction localized only in the physical region (the system to be studied). This assumption is needed to avoid the emergence of system-bath correlation effects which would complicate the overall formulation; 
further details can be found in Ref.~[\onlinecite{Ferretti2024PRB}].

A sketch of the embedding construction (discussed in detail in Sec.~\ref{sec:ham_embedding}) is provided in Fig.~\ref{fig:hamiltonian}:
working within the assumption discussed before of 
discrete energy levels and Green's function poles, 
we consider local dynamical external potentials of the form:
\begin{equation}
  \label{eq:local_embedding_main}
  v_{\text{ext}}(\mathbf{r},\omega) = 
     v_{0}(\mathbf{r}) + \sum_n \frac{R_n(\mathbf{r})}{\omega 
     -\Omega_n \pm i0^+},
\end{equation}
with $R_n(\mathbf{r}),\Omega_n \in \mathbb{R}$.
The above expression is a discrete sum over poles (see Refs.~\cite{Engel1991PRB,Savrasov2006prl,Chiarotti2022PRR,Chiarotti2023PhD,Chiarotti2024PRR,Ferretti2024PRB} for a  discussion of its analytical and algebraic properties in the context of Green's function theory).
In principle $\Omega_n$ could also be taken position dependent, but for simplicity (e.g., in order to avoid poles merging into continuous branch cuts) we do not consider this case here. Moreover, we will be interested in keeping the number of poles as small as possible, for physical transparency and computational reasons. 
In the embedding perspective mentioned above, the potential in Eq.~\eqref{eq:local_embedding_main} plays the role of an embedding self-energy. 
Importantly, the residues $R_n(\mathbf{r})$ must be taken positive in order to be interpreted as originating from embedding~\cite{Ferretti2024PRB,Martin-Reining-Ceperley2016book}, and, in general, for any self-energy to be well-behaved.
In fact, 
the positive-definiteness of the residues in Eq.~\eqref{eq:local_embedding_main} is a necessary condition~\cite{Stefanucci-vanLeeuwen2013book,Stefanucci2014PhysRevB,Ferretti2024PRB} for the resulting GF to have a positive-definite spectral function.

A formal presentation of the embedding construction leading to a local and dynamical external potential $v_\text{ext}(\mathbf{r},\omega)$ is presented in the following Section.

%
\subsection{Embedding formulation}
\label{sec:ham_embedding}
%
\begin{figure}
   \includegraphics[clip,width=0.45\textwidth]{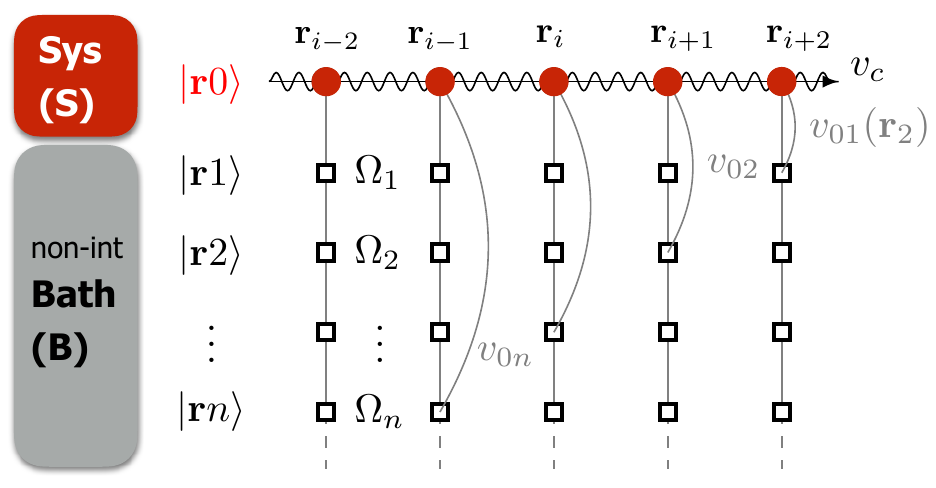}
   \caption{\label{fig:hamiltonian} (Color online) Cartoon representation of the extended Hamiltonian described in the text, showing how at every point $\mathbf{r}_i$ the system $S$ of interacting electrons (red) is embedded in a non-interacting bath $B$ (grey). Together, $S$ and $B$ form a closed system $C=S\cup B$. Both physical and auxiliary degrees of freedom are shown, together with their mutual coupling. The wiggly line indicates the degrees of freedom over which the Coulomb interaction $v_c$ is active.
   }
\end{figure}

Formally, the local dynamical external potential described in Sec.~\ref{sec:potential} can be obtained as an embedding potential for a class of Hamiltonians defined on an extended set of spacial degrees of freedom. 
Specifically, these are labeled by
$|\mathbf{r}n\rangle$, with $n=0,\dots N$, where $\mathbf{r}0$ (or simply $\mathbf{r}$) represents the three-dimensional physical space, while $n=1,\dots N$ provide additional degrees of freedom. Spin can be included by setting $| \mathbf{x}n\rangle \equiv | \mathbf{r}\sigma n\rangle$ and adopting the convention $\int d\mathbf{x} \to \sum_\sigma \int d\mathbf{r}$.

With this notation, we focus on the closed system $C=S\cup B$
that includes the interacting (physical) system $S$ spanned by $|\mathbf{x}0\rangle$, and a non-interacting bath $B$, described by $|\mathbf{x}n\rangle, \quad n>1$.
For this system, we define the following Hamiltonian:
\begin{eqnarray}
  \label{eq:full_hamiltonian}
  \bar{H} &=& \bar{T} + \bar{V}_{\text{ext}} + \bar{V}_{ee} 
   \\[7pt]  \nonumber
   \bar{T} &=& 
   \int \!d\mathbf{x} \, \hat{\psi}^\dagger(\mathbf{x}0) \, T_0(\mathbf{x})\, \hat{\psi}(\mathbf{x}0) 
   \\  \nonumber 
   \bar{V}_{\text{ext}} &=&  
   \int \!d\mathbf{x} \, \hat{\psi}^\dagger(\mathbf{x}0) \, v_0(\mathbf{x})\, 
   \hat{\psi}(\mathbf{x}0)
   \\  \nonumber
    &+&
   \sum_{n>0} \int \!d\mathbf{x} \,  
      \hat{\psi}^\dagger(\mathbf{x}n) \, \omega_n(\mathbf{x})\, \hat{\psi}(\mathbf{x}n)  \\ \nonumber
      &+& 
      \sum_{n>0} \int \!d\mathbf{x} \,  
      \hat{\psi}^\dagger(\mathbf{x}0) \, v_{0n}(\mathbf{x})\, \hat{\psi}(\mathbf{x}n)  + c.c.
   \\ \nonumber
   \bar{V}_{ee} &=& \frac{1}{2}\int\!d\mathbf{x}d\mathbf{x}'
   \, \hat{\psi}^\dagger(\mathbf{x}0) \hat{\psi}^\dagger(\mathbf{x}'0)\, v_c(\mathbf{x},\mathbf{x}') \, \hat{\psi}(\mathbf{x}'0) 
   \hat{\psi}(\mathbf{x}0) .
\end{eqnarray}
We use the bar symbol to indicate operators acting on the space of the closed system $C$. 
A cartoon describing the Hamiltonian structure 
is provided in Fig.~\ref{fig:hamiltonian}.
The first two terms represent the kinetic energy and the external potential, with the kinetic energy of the bath included in the latter in the on-site matrix element $\omega_n(\mathbf{x}) = T_n(\mathbf{x}) + v_n(\mathbf{x})$.
Notably, in the external potential, besides the usual local term $v_0(\mathbf{x})$ acting on the physical space, a local hopping contribution $v_{0n}(\mathbf{x})$ is added, coupling $\mathbf{x}0$ to $\mathbf{x}n$. These terms provide the system-bath coupling.
Importantly, the electron-electron interaction is taken to be nonzero only when acting on the physical coordinates, leaving the auxiliary bath non-interacting (this means that $\bar{V}_{ee}$ is just the physical interaction).
As discussed in Ref.~[\onlinecite{Ferretti2024PRB}], working with a non-interacting bath largely simplifies the treatment of interactions (see next Section).

As anticipated, in what follows 
all matrix elements are taken to be spin-independent, so that, e.g., $v_{n}(\mathbf{x})=v_{n}(\mathbf{r})$, $v_{0n}(\mathbf{x})=v_{0n}(\mathbf{r})$, etc.
Moreover, for simplicity we assume $v_{0n}(\mathbf{x}) = v_{n0}(\mathbf{x})  \in \mathbb{R}$, which is a sufficient condition to work with real valued wavefunctions and, in turn, real occupied Dyson orbitals $f_s(\mathbf{x}n) = \langle N-1,s| \hat{\psi}(\mathbf{x}n) | N,0\rangle$ on the whole $C$ system.
As already anticipated, for simplicity we also take the matrix elements $\omega_n$ (to become the poles of the embedding potential) to be independent of $\mathbf{r}$.

In the following we provide results concerning the embedding of the Green's function and of the Dyson orbitals for the systems introduced by Eq.~\eqref{eq:full_hamiltonian}, to be later used to define the total energies in Sec.~\ref{sec:total_energy}.
We conclude the Section providing a physical interpretation of the frequency dependence of the embedding potential $v_{\text{ext}}(\omega)$.

\subsubsection{Green's function embedding}
%
Given the Hamiltonian in Eq.~(\ref{eq:full_hamiltonian}) 
for the closed system $C=S\cup B$ (barred operators), 
the embedding potential~\cite{Stefanucci-vanLeeuwen2013book,Martin-Reining-Ceperley2016book,Ferretti2024PRB} acting on the physical space due to the coupling with the bath ($n\ge1$ components) becomes local and dynamical and has exactly the form of the external potential in Eq.~(\ref{eq:local_embedding_main}). In particular, one has 
\begin{eqnarray}
   \label{eq:embedding_defs1}
   R_n(\mathbf{r}) &=& \left| v_{0n}(\mathbf{r}) \right|^2, \\[5pt]
   \Omega_n 
   &=& T_n + v_{n} = \omega_n,
   \label{eq:embedding_defs2}
\end{eqnarray}
where, as already mentioned, $\omega_n$ is taken to be independent of $\mathbf{r}$.
Then, the Green's function $G$ for the embedded sub-system $S$ can be written as:
\begin{equation}
  G(\omega) = \Big[ \omega I -T - v_{\text{ext}}(\omega) -\Sigma_{\text{Hxc}}(\omega) \Big]^{-1},
  \label{eq:dyson_solution}
\end{equation}
where $\Sigma_{\text{Hxc}}(\omega)$ is the self-energy accounting for electron-electron interactions. 

As discussed in Ref.~\cite{Ferretti2024PRB}, while in principle $\Sigma_{\text{Hxc}}$ would depend on the GF of the whole closed system ($\bar{G}$), in view of the presence of interactions only in the physical region $S$, many-body perturbation theory only involves propagators in the physical system, making $\Sigma_{\text{Hxc}}=\Sigma_{\text{Hxc}}[G]$.
By considering discrete poles in $G$, the Dyson orbitals $|f_s\rangle$ appearing in Eq.~\eqref{eq:lehmann} obey also the equations:
\begin{eqnarray}
   \label{eq:dyson_qp}
   \Big[ T &+& v_{\text{ext}}(\epsilon_s) + \Sigma_{\text{Hxc}}(\epsilon_s) \Big] | f_s\rangle = \epsilon_s | f_s \rangle, \\[5pt]
   Z_s &=& \left[ 1 +\langle f_s | \Dot{v}_{\text{ext}}(\epsilon_s) + \Dot{\Sigma}_{\text{Hxc}}(\epsilon_s)| f_s \rangle \right],
   \label{eq:dyson_qp_Z}
\end{eqnarray}
where $\langle f_s | f_s \rangle = Z_s$, and we have defined 
\begin{equation}
  \label{eq:def_dot_v}
  \dot{v}_{\text{ext}}(\omega) = \frac{\partial v_\text{ext}(\omega)}{\partial \omega}, \qquad
  \dot{\Sigma}_{\text{Hxc}} = 
  \frac{\partial \Sigma_\text{Hxc}(\omega)}{\partial \omega}.
\end{equation}
An important consequence of working with discrete states is the possibility of dealing with a causal (e.g. time-ordered) dynamical potential in Eq.~\eqref{eq:local_embedding_main}, while keeping the potential Hermitian at the frequencies corresponding to the poles $\epsilon_s$ of the Green's function $G$. Since the external potential is also local, one has $v_{\text{ext}}(\mathbf{r},\epsilon_s) \in \mathbb{R}$. In fact, the non-Hermitian contributions to $v_\text{ext}$ are proportional to Dirac $\delta$'s centered around the poles $\Omega_n$ of the potential, which, by construction, are zeroes (and not poles) of $G$. 
See Appendix~\ref{sec:math_details} for further details.

\subsubsection{Dyson orbital embedding}

Alternatively to the GF formulation, one can work out the equation of motion (EOM) of the Dyson orbitals, Eqs.~(\ref{eq:amplitudes1}-\ref{eq:amplitudes2}), extended to the $\mathbf{r}n$-space, for the class of Hamiltonians in Eqs.~\eqref{eq:full_hamiltonian}. We adapt the derivation of Almbladh and von Barth~\cite{Almbladh1985PhysRevB} to the above Hamiltonians and obtain a set of coupled equations,
\begin{multline}
  \label{eq:dyson_r0}
  \left[-\frac{1}{2}\nabla^2 + v_0(\mathbf{r})\right] f_s(\mathbf{r}0) 
  + \sum_{n>0} v_{0n}(\mathbf{r}) f_s(\mathbf{r}n) + \\ 
  + \sum_{s'}^{\text{occ}} K_{ss'}(\mathbf{r}0) f_{s'}(\mathbf{r}0) = \epsilon_s f_s(\mathbf{r}0)
\end{multline}
\vspace{-12pt}
\begin{equation} 
    \label{eq:EOM_bath}
    \big[\epsilon_s -\Omega_n\big] 
    f_s(\mathbf{r}n) = v_{n0}(\mathbf{r}) f_s(\mathbf{r}0) 
    \qquad \qquad \qquad
\end{equation}
where
\begin{eqnarray}
  \label{eq:Kss_kernel}
  K_{ss'}(\mathbf{r}0) &=& \sum_\sigma\int d\mathbf{r}'  \, v_c(\mathbf{r},\mathbf{r}') \, 
  \times
  \\[5pt] 
  &\times&
        \langle N-1,s| 
  \hat{\psi}_\sigma^\dagger(\mathbf{r}' 0) \hat{\psi}_\sigma(\mathbf{r}'0) | N-1, s'\rangle .
  \nonumber
\end{eqnarray}
Notably, Eq.~\eqref{eq:EOM_bath} encodes the non-interacting nature of the bath. Moreover, in Eq.~\eqref{eq:Kss_kernel}, $K_{ss'}(\mathbf{r}0)$ depends on $\mathbf{r}$ and not on $\mathbf{x}$ (i.e. it does not depend on spin), because $v_c(\mathbf{x},\mathbf{x}')$ is spin-independent.

By eliminating the coupling terms from Eq.~(\ref{eq:dyson_r0}), one gets an EOM for the Dyson orbitals in the physical subsystem $f_s(\mathbf{r}0)$, where the external potential is replaced by the local dynamical potentials coming from embedding:
\begin{multline}
  \label{eq:dyson_r0_embed}
  \left[-\frac{1}{2}\nabla^2 + v_{\text{ext}}(\mathbf{r}0,\epsilon_s)\right] f_s(\mathbf{r}0) 
 \\ 
  +\sum_{s'}^{\text{occ}} K_{ss'}(\mathbf{r}0) f_{s'}(\mathbf{r}0) = \epsilon_s f_s(\mathbf{r}0)
\end{multline}
(from now on we'll drop the 0 index from $\mathbf{r}0$). The expression for $v_{\text{ext}}(\mathbf{r},\omega)$ is given by Eqs.~(\ref{eq:local_embedding_main}) and (\ref{eq:embedding_defs1}-\ref{eq:embedding_defs2}).
This equation for the Dyson orbitals is similar to the one
involving the self-energy operator [which can be referred to Dyson orbitals in the case of discrete poles, see, e.g., Eq.~(\ref{eq:dyson_qp})], except that it involves a spatially local term which is non-diagonal in the orbital indexes, while the self-energy is non-local but orbital-diagonal.

%
\subsubsection{Physical interpretation}
%
The frequency dependency of $v_{\text{ext}}(\omega)$ can be  interpreted
by writing the dynamical part of the external potential, seen as an embedding self-energy, in the time domain,
\begin{eqnarray}
  \nonumber
  v_{\text{ext}}(\mathbf{r},\tau) &=& v_0(\mathbf{r})\delta(\tau) + 
                                    \sum_n v_{0n}(\mathbf{r}) \, g_{n\mathbf{r}}(\tau) \, v_{n0}(\mathbf{r}), \\
            g_{n\mathbf{r}}(\tau) &=& -i \langle\, T\left[ c_n(\mathbf{r},\tau+t_0),c_n^\dagger(\mathbf{r},t_0) \right] \,\rangle,             
\end{eqnarray}
where 
and $g_{n\mathbf{r}}(\tau)$ 
is the time-ordered Green's function of the uncoupled bath (i.e. the bath when $v_{0n}(\mathbf{r})=0$) projected on mode $n\mathbf{r}$. 
In this view, the time dependence $\tau$ of the external potential comes from a dynamics internal to the bath system, where one particle is created/annihilated and evolved in time
for an interval $\tau$ 
(following a dynamics dictated by the poles $\Omega_n$'s). 
Such added/withdrawn particle is exchanged with the main system with a  
hopping probability proportional to $|v_{0n}(\mathbf{r})|^2$. At the same time, the main system also undergoes a dynamics with $N-1$/$N+1$ particles, which in turn probes its charged excitations.

\subsection{Total energies}
\label{sec:total_energy}
%
By taking advantage of the above results, in this Section we first consider the total energy for the closed system $C=S\cup B$ and discuss the quantities involved in the one-to-one correspondence of densities and potentials. 
Then, using the Klein functional, we identify a partition of the energy to be associated to the embedded system $S$ in order to allow for a variational determination of the problem.
By using the definitions in Eq.~\eqref{eq:full_hamiltonian} and writing $\bar{V}_\text{ext}$ in a more compact notation as 
\begin{equation}
   \bar{V}_{\text{ext}} = \sum_{nn'} \int \!d\mathbf{x} \,
         \hat{\psi}^\dagger(\mathbf{x}n) \, \bar{v}_{\text{ext}}(\mathbf{x}n,\mathbf{x}n')\, \hat{\psi}(\mathbf{x}n'),   
\end{equation}
the total energy and potential-response of the closed system $C$ 
can be written as:
\begin{eqnarray}
  \label{eq:etot_C}
  E_C[\bar{V}_{\text{ext}}] 
  &=&  
\langle N,0 | \bar{T} + \bar{V}_\text{ext} + \bar{V}_{ee} | N,0\rangle, 
  \\[7pt]
  \frac{\delta E_C}{\delta \bar{v}_{\text{ext}}(\mathbf{r}n,\mathbf{r}n')} &=& \bar{\gamma}(\mathbf{r}n',\mathbf{r}n), 
  \label{eq:potential_res}
  \\
  \bar{\gamma}(\mathbf{r}'n',\mathbf{r}n) &=& \sum_s^{\text{occ}} f_s(\mathbf{r}'n') f_s^{*}(\mathbf{r}n).
  \label{eq:potential_res2}
\end{eqnarray}
In the equations above we have exploited the Hellmann-Feynman theorem and $\bar{\gamma}$ is the one-body reduced density matrix of the whole $C$ system, in the extended space, corresponding to the ground state $| N,0\rangle$. The appearance of $\bar{\gamma}$ in Eq.~(\ref{eq:potential_res})  
is due to $\bar{V}_{\text{ext}}$ being local in $\mathbf{r}$, but non-local in $n,n'$.

Specializing Eqs.~\eqref{eq:potential_res} and~\eqref{eq:potential_res2} to the parameters defining $\bar{v}_\text{ext}$ and making use of Eq.~\eqref{eq:EOM_bath}, one obtains:
\begin{eqnarray}
     \label{eq:E_diff}
     \frac{\delta E_C}{\delta v_0(\mathbf{r})} &=& \sum_s^\text{occ} f_s(\mathbf{r}0)f_s^*(\mathbf{r}0) = \rho(\mathbf{r}) \\
     \nonumber
     \frac{\delta E_C}{\delta v_{0n}(\mathbf{r})} &=& \sum_s^\text{occ} f_s(\mathbf{r}n)f_s^*(\mathbf{r}0) = \sum_s^\text{occ} \frac{v_{n0}(\mathbf{r})\rho_s(\mathbf{r})}{\epsilon_s-\Omega_n}, \\
     \nonumber
     \frac{\delta E_C}{\delta\Omega_n(\mathbf{r})} &=& \sum_s^\text{occ} f_s(\mathbf{r}n)f_s^*(\mathbf{r}n) = \sum_s^\text{occ} \frac{R_{n}(\mathbf{r})\rho_s(\mathbf{r})}{(\epsilon_s-\Omega_n)^2},
\end{eqnarray}
where we have used Eq.~\eqref{eq:embedding_defs1} and the assumption $\Omega_n(\mathbf{r})=\Omega_n$. 
By writing the variation of the external dynamical potential of Eq.~\eqref{eq:local_embedding_main}, 
\begin{eqnarray}
  \label{eq:vext_variation}
  \delta v_\text{ext}(\mathbf{r},\omega) &=& \delta v_0(\mathbf{r}) 
         + \sum_n \left[ \frac{\delta R_n(\mathbf{r})}{\omega -\Omega_n} + 
         \frac{R_n(\mathbf{r})\delta\Omega_n}{(\omega-\Omega_n)^2}
         \right] \qquad
\end{eqnarray}
considering that $\delta R_n(\mathbf{r}) = \delta v_{0n}(\mathbf{r}) \, v_{n0}(\mathbf{r}) + v_{0n}(\mathbf{r}) \, \delta v_{n0}(\mathbf{r})$, and evaluating the complete variation of $E_C$ from Eqs.~\eqref{eq:E_diff}, one finally obtains:
\begin{eqnarray}
  \delta E_C[\bar{V}_{\text{ext}}] &=& \sum_s^{\text{occ}} \int d\mathbf{r}\, 
  \rho_s(\mathbf{r}) \, \delta v_\text{ext}(\mathbf{r},\epsilon_s) 
  \\ \nonumber
  &=& 
  \int d\mathbf{r} d\omega \, 
  \sum_s^{\text{occ}} \rho_s(\mathbf{r}) \delta(\omega-\epsilon_s) \, \delta v_\text{ext}(\mathbf{r},\omega) .
\end{eqnarray}
This is a first important result of the present work, showing that the local dynamical potential introduced with the embedding construction presented in Sec.~\ref{sec:ham_embedding} is actually conjugated to the 
spectral density.

%
%

We are now in the position to discuss the total energy of the embedded system. In order to do so we use the Green's function formalism and the Klein functional~\cite{Luttinger-Ward1960PR, Klein1961PR,Baym1961PhysRev,Almbladh1999InternationalJournalofModernPhysicsB} to express the total $E_C$ of Eq.~\eqref{eq:etot_C}. This functional depends on the GF of the whole system, $\bar{G}$, and is variational; i.e., its gradient is zero when evaluated using the physical one-particle GF of the problem. As derived and discussed in Ref.~[\onlinecite{Ferretti2024PRB}], in the presence of embedding the Klein functional can be partitioned in such a way to maintain variationality wrt the trial GF of the embedded system $G$.
The following expressions hold:
\begin{eqnarray}
     E^K_C[\bar{V}_\text{ext},\bar{G}] &=& E^{K}_S[v_{\text{ext}},G] + 
         E_{0B} \\[10pt]
     \label{eq:klein}
     E^{K}_S[v_{\text{ext}},G] 
                        &=&
                        \text{Tr}_\omega \left\{v_{\text{ext}}G\right\} + \Phi_{\text{Hxc}}[G]        
     \\[5pt] \nonumber
                         &+& \text{Tr}_\omega \left\{I -G_0^{-1}G \right\} 
                         -\text{Tr}_\omega \left\{v_0 G\right\} \\[4pt]
                         &+& \text{Tr}_\omega \left\{\text{Ln} \, G_0^{-1}G\right\} +
                        \text{Tr}_\omega \left\{h_0 G_0\right\} 
        \nonumber
     \\[7pt]   
     E_{0B} &=& \sum_n^{\text{occ}} \int_{D_n} d\mathbf{r} \, \Omega_n(\mathbf{r}),
     \nonumber 
     \\
     &=& \sum_n^{\text{occ}} V(D_n) \Omega_n.
     \label{eq:E0B}
\end{eqnarray}
Notably, the Tr operator in Eq.~\eqref{eq:klein} is intended to integrate only over the physical space $\mathbf{r}0$, i.e.~the $S$ subsystem.
In the expressions above, $h_0 = T + v_0$ and $G_0$ refer to an auxiliary (non-interacting and closed) reference system and do not contribute
to the actual value of the functional~\cite{Ismail-Beigi2010PhysRevB} (the dependency of $E^K$ on $G_0$ has been therefore dropped to simplify the notation).
As also discussed in Ref.~\eqref{eq:klein}[\onlinecite{Ferretti2024PRB}], since the bath $B$ is assumed non-interacting, the $\Phi_{\text{Hxc}}$ functional appearing in Eq.~\eqref{eq:klein} depends on the embedded Green's function $G$ and not on the whole $\bar{G}$ of the closed system $C$.
The term $E_{0B}$, involving the bare energies $\Omega_n$ of the extra degrees of freedom ($D_n$ being the domain of points where $\Omega_n$ are defined, and $V(D_n)$ its volume), is the energy of the uncoupled bath, i.e., when the $S$-$B$ hopping terms $v_{0n}(\mathbf{r})$ are set to zero. 
We also note that the region of space $D_n$ need therefore to be finite.
As mentioned above, this specific partition is needed to preserve the variationality of $E^K_S$ wrt $G$; more details are provided in Ref.~[\onlinecite{Ferretti2024PRB}].

With the above definitions, the energy contribution associated with the external dynamical potential 
can be evaluated using the Green's function $G$ of the system as
\begin{eqnarray}
   \!\!\! \nonumber
   \Delta E_{\text{ext}} 
           &=& 
        \int_{-\infty}^{+\infty} \! \frac{d\omega}{2 \pi i} \, e^{i\omega0^+} \!\!\! \int \!\! d\mathbf{x} \,
        v_{\text{ext}}(\mathbf{x},\omega) G(\mathbf{x},\mathbf{x}^+,\omega)
        \\ \nonumber
        [&=&] 
        \int_{-\infty}^\mu d\omega \int \! d\mathbf{r} \,\, 
        v_{\text{ext}}(\mathbf{r},\omega) \rho(\mathbf{r},\omega), \\ 
        &=& 
        \sum_s^\text{occ} \int \! d\mathbf{r} \,\,
        v_{\text{ext}}(\mathbf{r},\epsilon_s) \rho_s(\mathbf{r}), 
        \label{eq:vext_expectation}
\end{eqnarray}
yielding the first term in the rhs of Eq.~\eqref{eq:FT_partition}. 
As indicated by the $[=]$ sign,  
the equivalence of the first and last two lines in Eqs.~\eqref{eq:vext_expectation} is conditional. For it to to hold, one needs to be able to discard contributions from the poles of $v_{\text{ext}}(\omega)$ in the frequency integral. As a consequence of working with discrete poles, this is possible when $G$ and $v_{\text{ext}}$ are connected by a Dyson equation (i.e. when $G$ is computed in the presence of an embedding potential $v_{\text{ext}}$), since the poles of $v_{\text{ext}}$ then correspond to zeroes of $G$ (in the subspace spanned by each residue of $v_{\text{ext}}$);
see App.~\ref{sec:math_details} for a detailed discussion.

By taking into account that $\Sigma_{\text{Hxc}}[G] = 2 \pi i \frac{\delta \Phi_{\text{Hxc}}}{\delta G}$, 
the gradient of the Klein functional $E^K$ for the $S$ system (we have dropped the $S$ label for simplicity) is given by
\begin{equation}
  \label{eq:klein_gradient}
  2 \pi i \frac{\delta E^{K}}{\delta G} = v_{\text{ext}}(\omega) -v_0 + \Sigma_{\text{Hxc}}(\omega) -G_0^{-1} +G^{-1},\
\end{equation}
making $E^{K}_S$ stationary at the solution of the Dyson equation given in Eq.~(\ref{eq:dyson_solution}), i.e., the Green's function on subsystem $S$ from embedding theory. 
By exploiting the variationality of the Klein functional $E^K[v_\text{ext},G]$ wrt to $G$, one can express the derivative of the total energy wrt the external potential as
\begin{equation}
  \label{eq:klein_vext_variation}
  \frac{\delta E^K}{\delta v_\text{ext}(\mathbf{r},\omega)} = \frac{1}{2 \pi i} e^{i\omega 0^+} \sum_\sigma G_\sigma(\mathbf{r},\mathbf{r}^+,\omega),
\end{equation}
which, when evaluated for real variations of the potential (i.e. for frequencies far from its poles, while possibly close to the poles of $G$), makes the variation proportional to the spectral density $\rho(\mathbf{r},\omega)  =A(\mathbf{r},\mathbf{r},\omega)$.

In passing, we note that by using Eq.~\eqref{eq:vext_variation} jointly with Eq.~\eqref{eq:klein_vext_variation}, contributions to $\delta E^K$ from first order poles of $\delta v_{\text{ext}}$ are cancelled by zeros of $G$ (see App.~\ref{sec:math_details}), while second order poles (related to $\delta \Omega_n$) give a finite term. Eventually, this is then cancelled by the variation of $E_{0B}$ in Eq.~\eqref{eq:E0B}.
More precisely, one has
\begin{eqnarray}
   \delta E^K &=& \text{Tr}_\omega \left\{ \delta v_\text{ext}\, G\right\} 
       \\ \nonumber
       &=& \sum_s^{\text{occ}} \int d\mathbf{r} \, \rho_s(\mathbf{r}) \, \delta v_{\text{ext}}(\mathbf{r},\epsilon_s) + 
       \\ \nonumber
       &+& \sum_n^{\text{occ}} \text{Tr} \left\{ G(\omega) (\omega-\Omega_n)^{-1} \, R_n\right\}_{|\omega\to\Omega_n} \delta\Omega_n ,
\end{eqnarray}
where in writing the second term we have used $G(\Omega_n)=0$. The limit $G(\omega)(\omega-\Omega_n)^{-1}$ for $\omega \to \Omega_n$ can be evaluated according to Eq.~\eqref{eq:G_times_omega_minus_Omega} in App.~\ref{sec:math_details}, resulting in $-R_n^{-1}(\mathbf{r})$. 
Overall, this gives:
\begin{eqnarray}
  \delta E^K = \sum_s^{\text{occ}} \int d\mathbf{r} \, \rho_s(\mathbf{r}) \, \delta v_{\text{ext}}(\mathbf{r},\epsilon_s) - \sum_n^{\text{occ}} \int_{D_n} d\mathbf{r}  \, \delta\Omega_n,
  \nonumber
\end{eqnarray}
the last term being that canceling with the variation of Eq.~\eqref{eq:E0B}. 
%


\section{One-to-one correspondence}
\label{sec:one-to-one}
\begin{figure*}
   \centering
   \includegraphics[clip,width=0.75\textwidth]{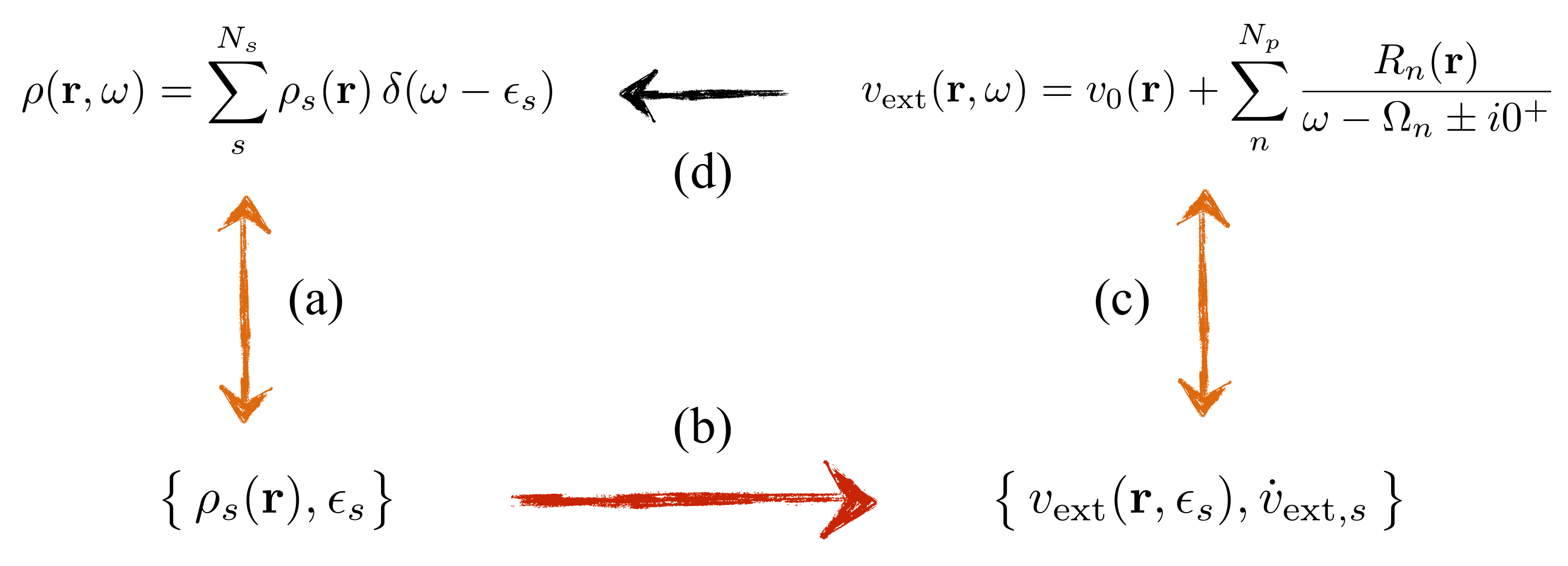}
   \caption{(Color online) Diagram describing the maps and the mathematical implications among the spectral density, external dynamical potential, and their representations. Here, $N_s$ and $N_p$ represent the number of occupied Dyson orbitals and of potential poles, respectively.
   Double arrows indicate a one-to-one mapping, single arrows a simple implication. Orange lines (a,c) sketch the equivalence of different representations (see Sec. \ref{sec:mapping}), while the black line (d) represents the direct implication potential-to-density. Importantly, the red thick line (b) indicates the inverse density-to-potential mapping, as proven in Sec.~\ref{sec:map_invertibility} and App.~\ref{sec:math_uniqueness_rational}.
   \label{fig:potential_density_mapping}
   }
\end{figure*}

As a first step in the formulation of \acronym{}, 
in this Section we derive a one-to-one correspondence,
$     \rho(\mathbf{r},\omega) 
      \longleftrightarrow v_{\text{ext}}(\mathbf{r},\omega), 
$
between occupied spectral densities and dynamical external potentials.
This is equivalent to answer the question whether a given occupied spectral density ($\omega < \mu$) is  generated by a unique (local and dynamical) external potential. 
In order to do so, first we discuss how to represent in a compact form spectral densities and potentials, allowing us to formulate the one-to-one correspondence, and then prove the invertibility of the map.

\subsection{Definition and representation of the potential-density map}
\label{sec:mapping}

Considering the definition of the spectral density given in Eq.~\eqref{eq:spectral_density}, when assuming discrete Dyson orbitals, one can 
write
\begin{equation}
  \rho(\mathbf{r},\omega) = \sum_s^{\text{occ}} \rho_s(\mathbf{r}) \, \delta(\omega-\epsilon_{s}),
\end{equation}
which leads to a representation of $\rho(\mathbf{r},\omega)$ in terms of the set of the  occupied Dyson orbital densities $\rho_s(\mathbf{r})$ and energies, i.e.,
\begin{equation}
   \rho(\mathbf{r},\omega) \longleftrightarrow 
   \, \big\{ \rho_s(\mathbf{r}), \epsilon_{s} \big\}. 
   \label{eq:repr_density}
\end{equation}
Importantly, we work with external dynamical potential, Eq.~\eqref{eq:local_embedding_main}, that are given in the form of a sum over poles~\cite{Chiarotti2022PRR,Chiarotti2023PhD,Chiarotti2024PRR,Ferretti2024PRB} (a frequency-independent part $v_0(\mathbf{r})$ plus poles $\Omega_n$ and residues $R_n(\mathbf{r})$), i.e.
\begin{equation}
   v_{\text{ext}}(\mathbf{r},\omega) \longleftrightarrow 
   \, \big\{ v_0(\mathbf{r}), \Omega_{n}, R_n(\mathbf{r}) \big\}. 
   \label{eq:repr_potental_poles}
\end{equation}
This representation highlights the analytical structure of the potential, rather than  
quantities of more direct use
such as, e.g., the values of the potential at the Dyson orbital energies. 

In view of this, it now becomes convenient to introduce an alternative representation of $v_{\text{ext}}(\mathbf{r},\omega)$ based on the theory of rational interpolation~\cite{Macon1962TheAmericanMathematicalMonthly,Antoulas1988LinearAlgebraanditsApplications,Graves-Morris1983NumerischeMathematik,Trefethen2019book_ch26}, where, based also on the the definitions in Eq.~\eqref{eq:def_dot_v}, we work with the potential and its space-averaged frequency derivative evaluated at a set of points $\omega_s$ (conditions discussed below): 
\begin{eqnarray}
   \label{eq:dynamical_sampling}
   v_{\text{ext},\,s}(\mathbf{r}) &=& v_{\text{ext}}(\mathbf{r},\omega_s) \\
   \dot{v}_{\text{ext},\,s}  &=& \int d\mathbf{r} \, \dot{v}_{\text{ext}}(\mathbf{r},\omega_s) \, p_s(\mathbf{r}),
   \nonumber
\end{eqnarray}
where $p_s(\mathbf{r})$ are some given  weights (these will later become the orbital densities $\rho_s(\mathbf{r})$).
Next, we note that the dynamical potential in Eq.~\eqref{eq:local_embedding_main}, as a function of frequency $\omega$, has the form of a rational function written as the ratio of two polynomials of order $N_p$, with $N_p$ being the number of poles in Eq.~\eqref{eq:local_embedding_main}. The frequency derivative of the potential is also a rational function, and one can therefore exploit results from rational interpolation~\cite{Macon1962TheAmericanMathematicalMonthly,Trefethen2019book_ch26}, granting the uniqueness of the underlying function $v_{\text{ext}}(\mathbf{r},\omega)$ once the sampling in Eq.~\eqref{eq:dynamical_sampling} is given.
A proof of this uniqueness, adapted to the present problem, is given in App.~\ref{sec:math_uniqueness_rational}.

For the present case, we set the sampling points and densities to be the $N_s$ occupied Dyson energies and densities (so, $\omega_s = \epsilon_s$ and $p_s(\mathbf{r})=\rho_s(\mathbf{r})$). Counting the degrees of freedom shows that,  
in order to determine the dynamical potential that generates the sampled values $\{v_{\text{ext},\,s}(\mathbf{r}),\,\dot{v}_{\text{ext},\,s}\}$,
one needs to have $N_p \leq N_s-1$; i.e., the number of poles needs to be smaller than the number of sampling points (here equal to that of Dyson orbitals) minus one.
This condition is
rather mild even for non-interacting systems, where the number of Dyson orbitals is significantly smaller than in the interacting case.

Given the uniqueness of the rational interpolation, one can formalize this alternative representation as:
\begin{equation}   
   v_{\text{ext}}(\mathbf{r},\omega)
   \longleftrightarrow  \big\{ v_{\text{ext}}(\mathbf{r},\epsilon_{s}), \dot{v}_{\text{ext},s} \big\}_{\rho}\;\; ,
   \label{eq:repr_potental_samples}
\end{equation}
where the subscript $\rho$ indicates that one needs to know the orbital energies and densities $\{\rho_s, \epsilon_s\}$ to define the expectation values $\dot{v}_{\text{ext},s}$ of $\dot{v}_\text{ext}(\mathbf{r},\epsilon_s)$.
Being based on a sampling of the dynamical potential, this alternative representation becomes very physical, since it involves the actual potentials $v_{\text{ext}}(\mathbf{r},\epsilon_{s})$ acting on the Dyson orbitals and the contribution $\Dot{v}_{\text{ext},s}$ of the dynamical external potential to the renormalization factors $Z_s$, according to Eqs.~(\ref{eq:dyson_qp},\ref{eq:dyson_qp_Z}).

Overall, the potential-density mappings resulting from these representations for the spectral density and the dynamical external potential are sketched in Fig.~\ref{fig:potential_density_mapping}. In particular: the black arrow indicates the direct potential-density map, stressing that the knowledge of the dynamical potential implies the knowledge of the spectral density via the direct solution of the electronic-structure problem;
Orange arrows mark instead the equivalence relations holding between the spectral density and its representation in terms of Dyson orbitals and energies, as well as between the dynamical potential and its representations in terms of poles or sampled values, respectively.
Finally, the thick red arrow indicates the reverse correspondence, the density-potential map, from the knowledge of the density to the dynamical potential that generates it.
The proof of the existence of this reverse correspondence, given in Sec.~\ref{sec:map_invertibility}, provides the invertibility of the potential-density map (one-to-one correspondence) and the equivalence of the knowledge of all quantities and representations sketched in Fig.~\ref{fig:potential_density_mapping}.
Notably, the knowledge of the occupied peaks of the spectral density is enough to determine the external potential over the whole frequency complex plane.

\subsection{Invertibility of the map}
\label{sec:map_invertibility_prep}
%

Mathematically, the invertibility condition of the map is equivalent to having a non-zero determinant for its Jacobian over the domain of dynamical potentials of interest. Using the representations in Eqs.~\eqref{eq:repr_density} and \eqref{eq:repr_potental_samples}, the Jacobian is a response function of the form
\begin{equation}
  \label{eq:spectral_chi_tot}
  \left(
   \begin{array}{c}
      \vdots \\
      \delta \rho_{s}(\mathbf{r}) \\
      \delta \epsilon_{s} \\
      \vdots
   \end{array}
   \right) 
   = \big( \chi \big)
   \left(
   \begin{array}{c}
       \vdots \\
      \delta v_{\text{ext}}(\mathbf{r}',\epsilon_{s'}) \\
      \delta \Dot{v}_{\text{ext},{s'}}  \\
      \vdots
   \end{array}
   \right) \;\; ,
\end{equation}
where the single components are
\begin{eqnarray}
  \label{eq:spectral_chi_1}
  \nonumber
  \chi_{11}(\mathbf{r},s; \mathbf{r}',s') &=& 
     \frac{\delta \rho_s(\mathbf{r})}{\delta v_{\text{ext}}(\mathbf{r}',\epsilon_{s'})}\; ,
     \\[4pt]  
  \nonumber 
  \chi_{12}(\mathbf{r},s; s') &=& 
     \frac{\delta \rho_s(\mathbf{r})}{\delta \dot{v}_{\text{ext},{s'}}}\; ,
     \label{eq:spectral_chi_2}
     \\[4pt]
  \nonumber
  \chi_{21}(s; \mathbf{r}',s') &=& 
     \frac{\delta \epsilon_s}{\delta v_{\text{ext}}(\mathbf{r}',s')}\; ,
     \\[4pt]
   \label{eq:spectral_chi}
   \chi_{22}(s; s') &=& 
   \frac{\delta \epsilon_s}{\delta \dot{v}_{\text{ext},{s'}}}\; .
\end{eqnarray}
One (or more) zero eigenvalues (i.e., a null  determinant) for the Jacobian correspond to a situation where $ \forall s, \ \delta \rho_s(\mathbf{r})$ and $\delta \epsilon_s$ can be zero for a non-trivial variation of the potentials. This entails that the spectral density could remain unchanged while varying the external potential from $v$ to $v+\delta v$, thereby breaking invertibility.

In order to address the invertibility of the potential-density map in the general case of interacting systems, we go back to the Hamiltonian $\bar{H}$ of the closed system containing both the physical system $S$ and the bath $B$, and in the presence of the static external potential $\bar{V}_{\text{ext}}$ (see Eq.~\eqref{eq:full_hamiltonian}). A change in the (static) external potential $\delta\bar{V}_{\text{ext}}$ induces a perturbation $\{\delta \rho_s(\mathbf{r}),\delta\epsilon_s\}$ in the spectral density. In order to demonstrate invertibility of the potential density map one has to show that if all $\delta \rho_s(\mathbf{r})$ and $\delta\epsilon_s$ are zero, then $\delta\bar{V}_{\text{ext}}$ needs also be zero.
   
Using the Hellmann-Feynman theorem, we write the variations in the charged excitation energies $(\epsilon_s = E^N_0 - E^{N-1}_s)$ as
\begin{eqnarray}
    \label{eq:delta_epsilon}
    \delta \epsilon_s &=& \delta E^N_0 -\delta E^{N-1}_s
    \\  \nonumber
    &=&  \langle N,0 | \delta\bar{V}_{\text{ext}} | N,0 \rangle - \langle N-1,s |  \delta\bar{V}_{\text{ext}} | N-1, s\rangle\; .
\end{eqnarray}
For any $s$, by setting $\delta \epsilon_s =0$, the two terms in the previous line sum to zero, so that $\langle N-1,s |  \delta\bar{V}_{\text{ext}} | N-1, s\rangle$ has to be at most equal to a $s$-independent constant.
For the orbital densities 
\begin{eqnarray}
  \label{eq:densities}
  \rho_s(\mathbf{r}) &=& |f_s(\mathbf{r})|^2 
  \\ \nonumber 
  &=&\langle N,0 | \hat{\psi}^\dagger(\mathbf{r}) | N-1,s \rangle \langle N-1,s | \hat{\psi}(\mathbf{r}) | N,0\rangle,
\end{eqnarray}
one needs to deal with the variations of both $|N,0\rangle$ and $|N-1,s\rangle$, leading to:
\begin{eqnarray}
  \delta \rho_s(\mathbf{r}) &=& f_s^*(\mathbf{r}) \, \langle N-1,s| \hat{\psi}(\mathbf{r}) | \delta N,0\rangle + c.c. 
  \nonumber \\
  &+& 
  f_s^*(\mathbf{r}) \, \langle \delta N-1,s| \hat{\psi}(\mathbf{r}) | N,0\rangle + c.c. \;,
  \label{eq:density_variations}
\end{eqnarray}
with $| \delta N,0\rangle$ and $| \delta N-1,s\rangle$ standing for variations of the $| N,0\rangle$ and $|N-1,s\rangle$ many-body states, respectively.
Using perturbation theory, and setting $\delta \rho_s(\mathbf{r})=0$, one obtains the following set of equations:
\begin{eqnarray}
   & &\sum_{m>0} f_s^*(\mathbf{r}) \, \langle N-1,s|\hat{\psi}(\mathbf{r}) | N,m\rangle 
   \frac{\langle N,m | \delta \bar{V}_{\text{ext}} | N,0\rangle }{E^N_{m} -E^N_0} 
   \nonumber \\ 
     &+&
  \sum_{p\neq s} f_s^*(\mathbf{r}) \, 
  \frac{\langle N-1,s | \delta \bar{V}_{\text{ext}} | N-1,p\rangle}{E^{N-1}_{p} -E^{N-1}_s} 
  \, f_p(\mathbf{r})
  \nonumber \\
   &+& c.c. =0  \qquad \qquad \qquad  \forall s,\mathbf{r} .
   \label{eq:delta_rhos_=0}
\end{eqnarray}
Equations (\ref{eq:delta_epsilon}) and (\ref{eq:delta_rhos_=0}) are the starting point for the invertibility proof in Sec.~\ref{sec:map_invertibility}.

\subsubsection{Regular and null Dyson orbitals}
\label{sec:regular_null_dyson}
%
Before proceeding further, it is important to explicitly discuss the occurrence of Dyson orbitals that are identically zero and do not contribute to the spectral density and the related mappings.
For instance, given the definitions in Eq.~\eqref{eq:amplitudes1}, or \eqref{eq:densities}, 
one can have Dyson orbitals of finite norm, that we refer to as {\it regular} ($R$),
but also orbitals that are zero everywhere, that we term {\it null} or {\it zero} ($Z$) Dyson orbitals.

As discussed below, null orbitals are expected to occur for ordinary systems and are not to be considered a trivial or exotic case.
For instance, null Dyson orbitals are obtained in non-interacting systems when a $|N-1,s\rangle$ state features multiple excitations wrt the ground state $|N,0\rangle$, or --- in the general case of spin-1/2 interacting Fermions --- when the $S_z$ spin component of $|N-1,s\rangle$ differs by more than 1/2 with that of $|N,0\rangle$.
{Similarly to the spin case, the presence of other symmetries could also lead to subspaces at $N-1$ particles that are not connected by a single field operator to the ground state $|N,0\rangle$, thereby leading to null orbitals.} Regular and null Dyson orbitals are further discussed in  App.~\ref{sec:regular_dyson_orbitals_dvext}.

This is relevant because when studying the variations of the spectral densities, it is important to distinguish the case of regular $R$ and null $Z$ Dyson orbitals when imposing $\delta \rho(\mathbf{r},\omega)=0$.
In particular, for regular orbitals $s$ one has  $\delta \epsilon_s =0$ while $\delta \rho_s(\mathbf{r})=0$ implies $\delta f_s(\mathbf{r})=0$, as from Eq.~\eqref{eq:density_variations}.
This is instead not true for null orbitals $t$, where $\delta f_t(\mathbf{r})$ and $\delta \epsilon_t$ are not necessarily zero. In fact, since $\delta \rho_t(\mathbf{r})= 0$ to linear order for null orbitals, $\delta \epsilon_t$ and $\delta f_t(\mathbf{r})$ corresponding to these orbitals do not contribute to $\delta \rho(\mathbf{r},\omega)$. Therefore, imposing the latter equal to zero does not have implications on $\delta \epsilon_t$ or $\delta f_t(\mathbf{r})$.
As discussed in Sec.~\ref{sec:proof3} and App.~\ref{sec:regular_dyson_orbitals_dvext}, this is important when proving the uniqueness of the external dynamical potentials.

\subsection{Proof of the invertibility} 
\label{sec:map_invertibility}
%
Having discussed the different ways to represent spectral densities and dynamical potentials, and set the stage concerning the potential-density mapping and its invertibility, in this Section we provide the formal result actually demonstrating the invertibility of the map. This is the first main result of the present work and can be stated as follows.
\begin{theorem}
Given a system of interacting electrons subject to a local dynamical external potential $v_{\text{ext}}(\mathbf{r},\omega)$, represented according to Eq.~\eqref{eq:repr_potental_poles}, and the corresponding occupied spectral density $\rho(\mathbf{r},\omega)$, represented according to Eq.~\eqref{eq:repr_density}, the map between the external potential and the occupied spectral density is invertible, provided that the number of poles $N_p$ in the potential is limited by the number of occupied Dyson orbitals $N_s$,
according to $N_p \leq N_s/2-1$. This condition can be further improved for non-interacting systems as $N_p \leq N_s-1$.
\label{theo:theorem-one}
\end{theorem}
{\it Proof.}
Following the discussion in the previous Sections, the proof of map invertibility is articulated across the following three steps: ($i$) First, exploiting the validity of Eq.~\eqref{eq:delta_rhos_=0} for all $s$ and $\mathbf{r}$, it is shown that the two terms in Eq.~\eqref{eq:density_variations} need to be separately zero. In turn, this proves that the $\langle N-1,s| \delta\bar{V}_{\text{ext}} | N-1,p\rangle$ matrix elements (with $s,p$ corresponding to regular Dyson orbitals; see Sec.~\ref{sec:regular_null_dyson}) are zero.
($ii$) Next, exploiting the above results together with Eqs.~(\ref{eq:EOM_bath},\ref{eq:dyson_r0_embed}) it is shown that $\delta v_{\text{ext}}(\mathbf{r},\epsilon_s)=0 $ for all (regular) $s$ and $\mathbf{r}$. ($iii$) Last, the uniqueness of the external dynamical potential is established.
These three steps are illustrated in the following.
Spin notation is omitted for simplicity.

\subsubsection{Variation of the spectral densities}
\label{sec:proof1}
%
We begin by imposing the conditions:
\begin{equation}
   \delta \rho_s(\mathbf{r}) = 0, \qquad \delta \epsilon_s = 0, \qquad \forall \mathbf{r},s .
   \label{eq:cond_delta_rhos}
\end{equation}
Nevertheless, rather than working directly with all $\delta \rho_s(\mathbf{r})$ according to Eq.~\eqref{eq:delta_rhos_=0}, we exploit the completeness of the $|N-1,s\rangle$ basis and introduce the auxiliary quantity $M_\alpha(\mathbf{r})$ defined as:
\begin{eqnarray}
   \label{eq:M_def}
M_\alpha(\mathbf{r}) &=& \sum_s^{\text{occ}} \rho_s(\mathbf{r}) \alpha(\epsilon_s), \\
  \nonumber
  &=& \langle N,0 | \psi^\dagger(\mathbf{r})\, \hat{\alpha}( \bar{F})  \, \psi(\mathbf{r}) | N,0\rangle,
\end{eqnarray}
where $\alpha(\omega)$ is a generic differentiable function (discussed later) and
we have defined:
\begin{equation}
   \bar{F}=E^N_0 \bar{I} -\bar{H}.
   \label{eq:def_A}
\end{equation}
Note that null Dyson orbitals do not contribute to Eq.~\eqref{eq:M_def} and that 
we use $\hat{\alpha}(\dots)$ when the function acts on operators. 
Notably, we have exploited $\epsilon_s = \langle N-1,s | \bar{F} | N-1,s\rangle$, to write
\begin{equation}
  \langle N-1,s | \,\hat{\alpha}(\bar{F}) \, | N-1,s' \rangle = \alpha(\epsilon_s) \delta_{s,s'}.
\end{equation}
The reason for introducing the function $\alpha(\omega)$ is to transfer the set of conditions 
in Eq.~\eqref{eq:cond_delta_rhos}
to the equivalent one
\begin{equation}
\delta M_\alpha(\mathbf{r})=0, \qquad\forall \mathbf{r}, \forall \alpha(\omega).
\end{equation}
Having restricted $s$ to regular Dyson orbitals ($s \in R$) does not pose problems since $\delta \rho_t(\mathbf{r})=0$ for null orbitals (see discussion in Sec.~\ref{sec:regular_null_dyson}).

The variation of $M_\alpha(\mathbf{r})$ can then be evaluated as 
$\delta M_\alpha(\mathbf{r}) = 
\delta T^{(1)}_{\alpha}(\mathbf{r}) + \delta T^{(2)}_{\alpha}(\mathbf{r})$, where
\begin{eqnarray}
  \label{eq_deltaM_alpha}
  \delta T^{(1)}_{\alpha}(\mathbf{r}) &=& \langle N,0 | \,\psi^\dagger(\mathbf{r}) \, \hat{\alpha}(\bar{F}) \,  \psi(\mathbf{r}) \, | \delta N,0\rangle + c.c. 
  \\[5pt]  \nonumber
  \delta T^{(2)}_{\alpha}(\mathbf{r}) &=& \langle N,0 | \,\psi^\dagger(\mathbf{r}) \, \delta\hat{\alpha}(\bar{F}) \,  \psi(\mathbf{r}) \, | N,0\rangle 
  \\[5pt] \nonumber
  &=& \sum_{ss'} f^*_s(\mathbf{r}) 
     \langle N-1,s| \, \delta \hat{\alpha}(\bar{F}) \, |N-1,s'\rangle
     f_{s'}(\mathbf{r})  .
\end{eqnarray}
In the perspective of looking at the variations of $M_\alpha(\mathbf{r})$, $\delta T^{(1)}_{\alpha}$ and $\delta T^{(2)}_{\alpha}$ encode the variations of the ground state and of the $|N-1,s\rangle$ excited states, respectively. As discussed in Sec.~\ref{sec:map_invertibility_prep}, our goal is to demonstrate that the two variations need to be zero independently.

Since we are working under the hypothesis that $\delta M_\alpha(\mathbf{r})=0$ for any $\alpha$, we can explore the consequences of this condition for certain given choices of the function $\alpha(\omega)$; in particular, we look for $\alpha(\omega)$ choices that make $\delta T^{(1)}_\alpha=0$. With this in mind, we start by considering functions that are non-zero in an arbitrary compact interval $]a,b[$ of the frequency axis, and are zero at the values of $\epsilon_s$ within this interval (with our mathematical setting of discrete states in periodic-boundary conditions there are only a finite number of $\epsilon_s$ in the chosen interval).  
One realization is
\begin{equation}
   \alpha(\omega) = g_1(\omega) \Theta(\omega) \, 
     \left[\prod_p^{n_p} (\omega -\epsilon_p)\right] \, \Theta(\omega) g_2(\omega),
     \label{eq:chosen_alpha}
\end{equation}
where $n_p$ is the number of poles $\epsilon_i$ in $]a,b[$, 
$\Theta(\omega)=\theta(\omega-a)\theta(b-\omega)$ is a product of Heaviside step functions, and $g_1(\omega)$, $g_2(\omega)$ are two arbitrary analytical functions.
In passing, $\Theta(\omega)$ 
is placed on both sides of Eq.~\eqref{eq:chosen_alpha} to simplify the next steps.
With the $\alpha(\omega)$ defined in Eq.~\eqref{eq:chosen_alpha},
one has $\delta T^{(1)}_{\alpha}(\mathbf{r})=0$ by construction, while $\delta T^{(2)}_{\alpha}(\mathbf{r})$ requires to evaluate $\delta \hat{\alpha}(\bar{F})$, that is:
\begin{multline}
   \langle N-1,s| \, \delta \hat{\alpha}(\bar{F}) \, |N-1,s'\rangle = g_1(\epsilon_s) \Theta(\epsilon_s) 
   \\[5pt] 
   \times \sum_{p=1}^{n_p} \prod_{i=1}^{p-1}(\epsilon_s-{\epsilon}_i) \, \langle N-1,s| \delta \bar{F} |N-1,s'\rangle \, 
   \\ 
   \times 
   \prod_{j=p+1}^{n_p} (\epsilon_{s'}-\epsilon_j) 
   \, \Theta(\epsilon_{s'}) g_2(\epsilon_{s'}),
\end{multline}
where, as one can readily verify, only the variation of the central product in Eq.~\eqref{eq:chosen_alpha} carries a non-zero contribution.
So, the hypothesis that $\delta M_\alpha(\mathbf{r})=0$, given that $\delta T^{(1)}_{\alpha}=0$ by construction, translates into having $\delta T^{(2)}_{\alpha}=0$, 
for any chosen $g_1(\omega)$ and $g_2(\omega)$. 
 
We consider an interval $]a,b[$ containing both $\epsilon_s$ and $\epsilon_s'$, and differentiate $\delta T^{(2)}_{\alpha}$ (last line of Eq.~51, and using the matrix elements from Eq.~53) with respect to $g_1(\epsilon_s)$ and $g_2(\epsilon_{s'})$, obtaining:
\begin{equation}
   f^*_s(\mathbf{r})\, \langle N-1,s| \delta \bar{F} |N-1,s'\rangle \, f_{s'}(\mathbf{r}) =0 \qquad \forall s,s' \; .
   \label{eq:key_zero_mat_element1}
\end{equation}
According to Eq.~\eqref{eq:def_A}, $\delta \bar{F}_{ss'} = 
\langle N-1,s |  \delta\bar{F}| N-1, s'\rangle=
-\langle N-1,s |  \delta\bar{V}_{\text{ext}} | N-1, s'\rangle=
-\delta \bar{V}_{\text{ext},ss'}$ for $s\neq s'$, and $\delta \bar{F}_{ss}=\delta \epsilon_s$ for for $s=s'$.
Importantly, we note that for any pair of regular (i.e., not identically zero) Dyson orbitals $f_s(\mathbf{r})$ and $f_{s'}(\mathbf{r})$, 
one has
from Eq.~\eqref{eq:key_zero_mat_element1} that
\begin{equation}
   \label{eq:key_zero_mat_element2}
   \delta \bar{V}_{\text{ext},ss'} =
   \langle N-1, s | \delta \bar{V}_{\text{ext}} | N-1, s' \rangle =0 \qquad \forall s,s' \in R.
\end{equation}
This is a key intermediate result that completes the first step $(i)$ of the proof of Theorem~\ref{theo:theorem-one}.

In passing, Eq.~\eqref{eq:key_zero_mat_element2} together with Eq.~\eqref{eq:delta_rhos_=0} show that for each regular Dyson orbital $s$
\begin{eqnarray}
   & &\sum_{m>0} f_s^*(\mathbf{r}) \, \langle N-1,s|\hat{\psi}(\mathbf{r}) | N,m\rangle 
   \frac{\langle N,m | \delta \bar{V}_{\text{ext}} | N,0\rangle }{E^N_{m} -E^N_0} 
   \nonumber \\ 
   & & 
   + c.c. =0,  
\end{eqnarray}
where the variation $|\delta N,0\rangle$ is  
decoupled from the variations $|\delta\, N-1,s\rangle$.

\subsubsection{Variation of the external embedding potential}
\label{sec:proof2}
%
We are now in the position of exploiting Eqs.~\eqref{eq:key_zero_mat_element1} and \eqref{eq:key_zero_mat_element2}, together with the equation of motion (EOM) for the Dyson orbitals, Eqs.~(\ref{eq:EOM_bath},\ref{eq:dyson_r0_embed}), to show that $\delta v_{\text{ext}}(\mathbf{r},\epsilon_s)=0$ for each regular orbital $s$.

We start by considering Eq.~\eqref{eq:dyson_r0_embed} for a regular orbital $s$, and calculate its variations to linear order. We impose $\delta f_{s'}(\mathbf{r})=\delta \epsilon_{s'}=0$ for all regular orbitals (this would not be the case for null orbitals $t$, for which $\delta f_t(\mathbf{r})$ may be non zero, as discussed in Sec.~\ref{sec:regular_null_dyson}).
This leads to the equation:
\begin{eqnarray}
   \delta v_\text{ext}(\mathbf{r},\epsilon_s) f_s(\mathbf{r}) 
   &=& 
   -\sum^R_{s'} \delta K_{ss'}(\mathbf{r}) f_{s'}(\mathbf{r}) 
   \nonumber \\
   & & -\sum^Z_{t} K_{st}(\mathbf{r}) \, \delta f_{t}(\mathbf{r}),
   \label{eq:dvext_eom}
\end{eqnarray}
where the kernel $K_{ss'}(\mathbf{r})$ is defined in Eq.~\eqref{eq:Kss_kernel}.

If we were to disregard the existence of null Dyson orbitals (i.e., assume that the $Z$ set is empty and all orbitals are regular), only the first term of the rhs of Eq.~\eqref{eq:dvext_eom} would have to be considered.
Then, as shown in App.~\ref{sec:regular_dyson_orbitals_dvext}, $\delta K_{ss'}(\mathbf{r})$ can be explicitly expressed --- using perturbation theory --- in terms of the matrix elements $\delta \bar{V}_{\text{ext},ss'}=\langle N-1,s | \delta \bar{V}_{\text{ext}} | N-1, s'\rangle$; for regular orbitals these are zero (Eq.~\eqref{eq:key_zero_mat_element2}), leading to the following relation:
\begin{equation}
\label{eq:dvext_=zero}
   \delta v_\text{ext}(\mathbf{r},\epsilon_s) 
   = 0 \qquad \qquad \forall s\in R.
\end{equation}
This is the second intermediate key result in the proof, 
showing that for any arbitrary variation of the external potential $\bar{V}_{\text{ext}}$ the variations of the local embedding potentials $v_\text{ext}(\mathbf{r},\epsilon_s)$ acting on regular Dyson orbitals need also to be zero.

The discussion of Eq.~\eqref{eq:dvext_eom} in the presence of both regular and null Dyson orbitals is presented in App.~\ref{sec:regular_dyson_orbitals_dvext}, where, at least for the cases of null orbitals originating from, e.g., spin symmetries, the rhs of Eq.~\eqref{eq:dvext_eom} is also proven to be zero.
%

\subsubsection{Uniqueness of the dynamical external potential}
\label{sec:proof3}
%
In this section we complete the proof of Theorem~\ref{theo:theorem-one}; starting from Eq.~\eqref{eq:dvext_=zero},  we show that
the $v_\text{ext}(\mathbf{r},\omega) \rightarrow \rho(\mathbf{r},\omega) $ map is invertible, if $v_\text{ext}(\mathbf{r},\omega)$ originates from a finite 
(discussed later) number of poles.

We begin by discussing the case of non-interacting electrons. With this hypothesis, in addition to having $\delta v_\text{ext}(\mathbf{r},\epsilon_s) =0$ for each regular orbital, as specified by Eq.~\eqref{eq:dvext_=zero}, one has also that Eq.~\eqref{eq:dyson_qp_Z} simplifies to
\begin{equation}
Z_s\,=\,1 +\int d\mathbf{r}\, \rho_s(\mathbf{r}) \,\Dot{v}_{\text{ext}}(\mathbf{r},\epsilon_s)\;.
\end{equation}
Imposing the conservation of the norm $Z_s$ of the regular Dyson orbitals by setting $\delta Z_s=0$ --- and recalling the hypothesis that $\delta \rho_s(\mathbf{r})=0$ --- leads to
\begin{equation}
   \label{eq:ddotvext_=zero}
   \int d\mathbf{r} \rho_s(\mathbf{r}) \,\delta \dot{v}_\text{ext}(\mathbf{r},\epsilon_s)\;=\;\delta \Dot{v}_{\text{ext},\,s}\;=\;0 \qquad \forall s\in R.
\end{equation}
So, we have shown that, having $\forall s$ $\{\delta \rho_s(\mathbf{r}) = 0,\delta\epsilon_s=0\}$ implies that 
$\{\delta v_\text{ext}(\mathbf{r},\epsilon_s)=0,\delta \Dot{v}_{\text{ext},\,s}=0\}$.
According to Sec.~\ref{sec:map_invertibility_prep}, 
this completes the proof of the potential-density map invertibility in the non-interacting case.

Notably, the reconstruction of the whole external potential in terms of $\big\{ v_0(\mathbf{r}), \Omega_{n}, R_n(\mathbf{r}) \big\}$ from the knowledge of $\big\{ v_{\text{ext}}(\mathbf{r},\epsilon_s), \dot{v}_{\text{ext},s} \big\}$ requires a constraint on the number of poles of the potential $N_p$, which needs to be smaller than the number of regular Dyson orbitals $N_s$, i.e. $N_p \leq N_s -1$ (see Sec.~\ref{sec:map_invertibility}). This condition defines the limit on the number of poles generating the external potential to have the map invertibility.

Next, we consider the case of interacting electrons. Since the number of regular orbitals in the presence of interactions is typically much larger, we avoid to make reference to $\delta \dot{v}_{\text{ext},\,s}$ as in Eq.~\eqref{eq:ddotvext_=zero}, and will just use Eq.~\eqref{eq:dvext_=zero}. This will lead to a different (weaker) constraint on the number of poles, in itself less relevant given this larger number of regular orbitals. We first note that the external potential $v_\text{ext}(\mathbf{r},\omega)$ is by construction a sum-over-poles, i.e., a rational function of the frequency (see Eq.~\eqref{eq:local_embedding_main}). The same holds for its variation, Eq.~\eqref{eq:vext_variation}. Thus, one can use the uniqueness results for rational interpolation, as detailed in App.~\ref{sec:math_uniqueness_rational}, adapted to the present case. 
Specifically,
according to Eq.~\eqref{eq:vext_variation}, $ \delta v_\text{ext}(\mathbf{r},\epsilon_s)=0$ can be written as:
\begin{eqnarray}
  \label{eq:dvext_repr_poles} 
  \delta v_0(\mathbf{r}) + \sum_n^{N_p} \frac{\delta R_n(\mathbf{r})}{(\epsilon_s-\Omega_n)} + \sum_n^{N_p} \frac{R(\mathbf{r})\delta\Omega_n}{(\epsilon_s-\Omega_n)^2} = 0,
\end{eqnarray}
where $N_p$ is the number of poles. 
It is important to recall that the representation of $v_\text{ext}(\mathbf{r},\omega)$, e.g., by means of $\big\{ v_0(\mathbf{r}), \Omega_{n}, R_n(\mathbf{r}) \big\}$ according to Eq.~\eqref{eq:repr_potental_poles}, is known; we want to find $\big\{\delta v_0(\mathbf{r}), \delta \Omega_n, \delta R_n(\mathbf{r}) \big\}$ and prove that they need to be zero. This in turn will imply that $\delta v_\text{ext}(\mathbf{r},\omega) = 0 $, as a consequence of having imposed $\delta \rho(\mathbf{r},\omega) =0$, thereby concluding the proof of Theorem~\ref{theo:theorem-one}.

Technically, we first observe that $\delta v_\text{ext}(\mathbf{r},\omega)$ in Eq.~\eqref{eq:repr_potental_poles} features first and second order poles, and   
can be written, as a function fo the frequency, as the ratio of two polynomials of degree $2N_p$, 
\begin{equation}
   \label{eq:dvext_poly_ratio}
  \delta v_\text{ext}(\mathbf{r},\omega) = \frac{\sum_{m=0}^{2N_p} A_m(\mathbf{r}) \,\omega^m}{\prod_{n=1}^{N_p} (\omega-\Omega_n)^2 },
\end{equation}
where the denominator is known and the unknowns $\big\{\delta v_0(\mathbf{r}), \delta \Omega_n, \delta R_n(\mathbf{r}) \big\}$ are included in $\{A_m(\mathbf{r})\}$.
Imposing $ \delta v_\text{ext}(\mathbf{r},\epsilon_s)=0, \quad \forall s=1,N_s$ from 
Eq.~\eqref{eq:dvext_repr_poles} is then equivalent to
\begin{equation}
   \sum_{m=0}^{2N_p} A_m(\mathbf{r}) \,\epsilon_s^m = 0.
\end{equation}
If the number $N_s$ of sampling frequencies $\epsilon_s$ corresponding to regular orbitals is at least $2 N_p+1$, the polynomial with $A_m(\mathbf{r})$ coefficients in the numerator of Eq.~\eqref{eq:dvext_poly_ratio} must be identically zero, since it has a number of roots larger than its degree ($2 N_p$). Having proven than $A_m(\mathbf{r})=0, \quad \forall m=0..2N_p$ in turn implies that  $\big\{\delta v_0(\mathbf{r}), \delta \Omega_n, \delta R_n(\mathbf{r}) \big\}\,=\,0$, which is the thesis.

In summary, the condition $N_p \leq N_s/2 -1$ on the number of poles grants the invertibility of the potential-density map in the interacting case.
We note that, despite the condition on $N_p$ being more strict than for the non-interacting case, this is typically not a problem for interacting systems in view of the very large number of regular Dyson orbitals.

This completes the proof of Theorem~\ref{theo:theorem-one}.
$\Box$


\section{Universal functional and total energy}
\label{sec:SF-universal}

\subsection{Functional definition and variational properties}
\label{sec:SF-universal-def}
In analogy to density-functional theory, where the one-to-one mapping between (static) densities and potentials allows one to define a universal functional of the density in order to provide a variational principle for the total energy, here, exploiting the one-to-one mapping proven in Sec.~\ref{sec:one-to-one}, it becomes possible to define a universal
functional of the occupied spectral density. Moreover, by adding the contribution
of an external dynamical potential, and taking into account the total energy expression given by the Klein functional~\cite{Luttinger-Ward1960PR, Klein1961PR,Baym1961PhysRev,Almbladh1999InternationalJournalofModernPhysicsB} in Eq.~(\ref{eq:klein}), a total energy functional of the occupied spectral density and a stationarity condition are provided. This will be the second main result of the present work, following the one-to-one correspondence demonstrated earlier.

First, we consider the set of occupied spectral densities $S_v$ that are interacting $v$-representable; i.e., that correspond, given a form for the interaction (Coulombic, typically), to GFs obtained
from a local and dynamical external potential as defined in Eq.~\eqref{eq:local_embedding_main}. 
In particular, in view of the one-to-one correspondence, for each $\rho(\mathbf{r},\omega) \in S_v$, there exists a unique $v_\rho(\mathbf{r},\omega)$, and thus a unique $G_\rho$ from the solution of the many-body problem; this unique $G_\rho$ has $\rho(\mathbf{r},\omega)$ as occupied spectral density, and makes the Klein functional $E^{\text{K}}[v_\rho,G]$ stationary with respect to variations in $G$. Following Eq.~\eqref{eq:klein_gradient}, we have
\begin{eqnarray}
   G_\rho(\omega) &=& \Big[\omega I -T -v_{\rho}(\omega) 
   -2\pi i\frac{\delta \Phi_{\text{Hxc}}}{\delta G_\rho}\Big]^{-1} ,
\end{eqnarray}
where $\rho(\mathbf{r},\omega)$ is 
$\frac{1}{\pi} \sum_\sigma \text{Im} [G_\sigma(\omega)]_{\mathbf{r},\mathbf{r}}$, 
as in Eq.~\eqref{eq:G_imag}.

Therefore, 
we can define a universal functional $F^{\text{SF}}[\rho]$ of the occupied spectral density as
\begin{eqnarray}
    \label{eq:HKdef_F_SF}
     F^{\text{SF}}[\rho] &=& 
     \Phi_{\text{Hxc}}[G_\rho]        
     \\ \nonumber
                         &+& \text{Tr}_\omega \left\{I -G_0^{-1}G_\rho \right\} 
                         -\text{Tr}_\omega \left\{v_0 G_\rho\right\} 
                         \\[4pt] \nonumber
                         &+& \text{Tr}_\omega \left\{\text{Ln} \, G_0^{-1}G_\rho\right\} +
                        \text{Tr}_\omega \left\{h_0 G_0\right\},
\end{eqnarray}
where, as for the Klein functional in Eq.~\eqref{eq:klein}, $G_0$ is an auxiliary  GF corresponding to an arbitrary non-interacting Hamiltonian $h_0 = T+ v_0$, used to write the functional but not contributing to it~\cite{Ismail-Beigi2010PhysRevB}. For later use, we call the five terms in Eq.~\eqref{eq:HKdef_F_SF} as $F_1,..,F_5$, respectively.

At this point, for any local external potential $v_{\text{ext}}(\mathbf{r},\omega)$ and for any $v$-representable occupied spectral density $\rho(\mathbf{r},\omega) \in S_v$, we can define the total energy functional
\begin{eqnarray}
   \label{eq:HKdef_SF}
   E^{\text{SF}}[v_\text{ext},\rho] &=& E^{\text{K}}[v_{\text{ext}},G_\rho] \\ 
        &=& \text{Tr}_\omega \left\{v_{\text{ext}} G_\rho \right\}
        + F^{\text{SF}}[\rho]\; .
    \nonumber
\end{eqnarray}
Parallel with the Hohenberg-Kohn (HK) case~\cite{Hohenberg-Kohn1964PR,Eschrig2003book} for the (static) charge density, we have written here a total energy functional $E^{\text{SF}}$ 
as the
sum of the universal functional of the spectral density $F^{\text{SF}}[\rho]$ plus the contribution of 
an external dynamical potential~\cite{note_klein_G0}. 
When $\rho(\mathbf{r},\omega)$ is the spectral density corresponding to the external potential 
$v_{\text{ext}}(\mathbf{r},\omega)$, $E^{\text{SF}}$ becomes the physical total energy given by the Klein functional.

For the $E^{\text{SF}}$ functional of Eq.~\eqref{eq:HKdef_SF}, the functional derivative wrt variations of the spectral density $\rho(\mathbf{r},\omega)$ can be written as:
\begin{equation}
   \label{eq:stationarity}
   \frac{\delta E^{\text{SF}}[v_\text{ext},\rho]}{\delta \rho} = 
       \frac{\delta E^{\text{K}}[v_\text{ext},G_\rho]}{\delta G_\rho}\,
       \frac{\delta G_\rho}{\delta \rho}.
\end{equation}
Centrally, the above expression becomes stationary when evaluated at the spectral density $\rho$ corresponding to the external potential $v_{\text{ext}}(\mathbf{r},\omega)$, thanks to the stationarity of $\frac{\delta E^{\text{K}}}{\delta G_\rho}$ when evaluated at the unique $G_\rho$ corresponding to $\rho$ (importantly, this property would not be valid when using the Galitskii-Migdal expression for the total energy).

Moreover, when the trial spectral density $\rho(\mathbf{r},\omega)$ in the functional of Eq.~\eqref{eq:HKdef_SF} is the physical spectral density corresponding to $v_\text{ext}(\mathbf{r},\omega)$ and $G_\rho$ becomes the physical GF, according to Eq.~\eqref{eq:vext_expectation} one also has:
\begin{eqnarray}
  \label{eq:exp_stationarity}
  \text{Tr}_\omega \left\{v_{\text{ext}} G_\rho \right\} &=&
  \sum_s^{\text{occ}} \int d\mathbf{r}d\omega \, v_{\text{ext}}(\mathbf{r},\epsilon_s) \rho_s(\mathbf{r}) 
  \\ \nonumber
  &=&   
  \int d\mathbf{r}d\omega \, v_{\text{ext}}(\mathbf{r},\omega) \rho(\mathbf{r},\omega).
\end{eqnarray} 
As discussed in Appendix~\ref{sec:math_details}, this is due to the fact that $v_\text{ext}(\mathbf{r},\omega)$ appears in the Dyson equation for $G_\rho$, Eq. \eqref{eq:dyson_solution}, so that its poles become zeros of $G_\rho$ and do not contribute to the expression of 
$\text{Tr}_\omega \left\{v_{\text{ext}} G_\rho \right\}$.
Working with dynamical potentials, Eq.~\eqref{eq:exp_stationarity} is nevertheless not valid for a generic Green's function $G_\rho$ (see below).

It is therefore tempting to introduce a second spectral functional, where the contribution from the external potential does not involve $G_\rho$ and is in the form of the second line of Eq.~\eqref{eq:exp_stationarity}:
\begin{eqnarray}
   \label{eq:HKdef_SF_var}
   \widetilde{E}^{\text{SF}}[v_{\text{ext}},\rho] = \int \!\! d\mathbf{r}d\omega \, v_{\text{ext}} 
   \rho 
   + F^{\text{SF}}[\rho].
\end{eqnarray}
This functional provides a more effective and natural way of 
expressing the contribution of the external potential, 
while assuming the same value of ${E}^{\text{SF}}$ in Eq.~\eqref{eq:HKdef_SF}
at the physical $\rho(\mathbf{r},\omega)$ corresponding to $v_\text{ext}(\mathbf{r},\omega)$. Generally, this comes at the price of losing the stationarity condition of Eq.~\eqref{eq:stationarity} at the physical solution.

Nevertheless, if we restrict the functional domain to the set of spectral densities originating from $v_\rho(\omega)$'s having (at least) the same occupied poles of $v_\text{ext}(\omega)$, Eq.~\eqref{eq:exp_stationarity} is identically valid, and variationality is restored.
First, we note that the restricted domain defined above is non trivial (i.e., non empty): as an example, one can build a set of spectral densities originating from dynamical potentials as in Eq.~\eqref{eq:local_embedding_main} by modifying the parameters $v_0(\mathbf{r})$ and $R_n(\mathbf{r})$, and/or by adding more poles.
Second, this restricted domain is the one that will be exploited in the formulation of a dynamical Kohn-Sham-like mapping (see Sec.~\ref{sec:KS-mapping} below).

As mentioned in Sec.~\ref{sec:spectral_density}, within the DMFT perspective, Savrasov and Kotliar developed~\cite{Savrasov2004PhysRevB} a functional theory of the local Green's function $G_\text{loc}(\mathbf{r},\mathbf{r}',\omega)$ with $|\mathbf{r}-\mathbf{r}'| \leq d_\text{cut}$; this involves both occupied and empty states, while in the present work we focus on the occupied spectral density $\rho(\mathbf{r},\omega)$. In passing, we also note that their functional formulation exploits a Legendre transform (that, strictly speaking, would require a convexity condition), as can also be done for DFT~\cite{Eschrig2003book}, to go from a functional of the external local field to a functional of the local Green's function. While in Eq.~\eqref{eq:HKdef_SF} we have followed the Hohenberg and Kohn construction~\cite{Hohenberg-Kohn1964PR,Eschrig2003book}, in view of the map invertibility proven in Sec.~\ref{sec:one-to-one}, a Legendre transform approach could also be used here.
%

\subsection{Constrained search formulation}
%
In this Section we provide an alternative formulation of the SF functional based on a constrained search formulation.
The definition of the SF functional presented above makes reference to the set $S_v$ of $v$-representable spectral densities, and parallels the Hohenberg-Kohn construction~\cite{Hohenberg-Kohn1964PR,Eschrig2003book} of DFT.
In the context of DFT, by means of the Levy-Lieb constrained search (CS)~\cite{Levy1979PNAS,Lieb1983IntJQuantumChem} the domain $S^n_v$ of the HK functional can be enlarged to the set $S^n_N$ of $N$-representable densities (i.e., densities $n(\mathbf{r})$ that can be obtained integrating a proper many-body wavefunction).
In practice, a constrained search over all possible wavefunctions giving the target density $n(\mathbf{r})$ is performed, resulting in the DFT formulation:
\begin{eqnarray}
    F^{\text{DFT,CS}}[n] &=& \text{min}_{|\Psi \rangle \rightarrow n} \, \langle \Psi | T + V_{\text{ee}} | \Psi \rangle,
    \\[5pt]
  \nonumber
    E^{\text{DFT,CS}}[v_{\text{ext}},n] 
       &=& \int \!\!d\mathbf{r}\, v_{\text{ext}}(\mathbf{r}) n(\mathbf{r}) + F^{\text{DFT,CS}}[n],
  \label{eq:levy}
\end{eqnarray}
which represents a generalization of the HK scheme since $S^n_v \subset S^n_N$ 
(the HK and CS functionals being the same for densities in $S^n_v$).
Notably, the variational (minimum) principle of the total energy as a functional of the many-body wavefunction has been used.

In order to build a similar construction for the SF functional, we first note that the occupied spectral density is a contraction of the one-particle GF $G$ rather than of a many-body wavefunction.
In this respect, the Klein functional~\cite{Luttinger-Ward1960PR, Klein1961PR,Baym1961PhysRev,Almbladh1999InternationalJournalofModernPhysicsB}, already used in the SF definition of Eq.~\eqref{eq:HKdef_SF}, provides an exact total energy expression as a function of $G$ and is equipped with a variational principle (in the form of a stationarity rather than a minimum condition, at variance with the wave-function based case).

In order to expand the set of $v$-representable spectral densities, we introduce the notion of $G$-representability as the set $S_G$ 
of spectral densities $\rho(\mathbf{r},\omega)$ that can be obtained from a well-behaved~\cite{Ferretti2024PRB} Green's function $G_\rho$ (featuring positive semi-definite residues and real poles, but not necessarily corresponding to a local and dynamical external potential).
Since  
$S_v \subset S_G$,
moving to $S_G$ implies that the one-to-one mapping $G_\rho  \longleftrightarrow \rho$ is not granted in general. Nevertheless, one can invoke (see below) a variational principle, as in Eq.~\eqref{eq:levy}, to overcome this difficulty and define a $F^{\text{SF}}$ functional.

In particular, following Ref.~\cite{Ismail-Beigi2017JPhysCondensMatter}, and taking $w=w(\mathbf{r},\omega)$ to be a local and dynamical potential as in Eq.~\eqref{eq:local_embedding_main}, and $G$ any well-behaved GF (i.e., having real poles and positive semi-definite Hermitian residues~\cite{Ferretti2024PRB}), we introduce the additional functional $D[w,G]$ that represents the norm of the gradient of the Klein functional $E^K[w,G]$ (which is minimum and not just stationary at the physical $G$, solution of the problem defined by $w$).
Explicitly, we can write:
\begin{eqnarray}
   \label{eq:grad_functional1}
   D[w,G] &=& \left\| \,2\pi i\frac{\delta E_{\text{K}}[w,G]}{\delta G}\, \right\|^2  \\
   \nonumber
             &=& 
   \left\| \omega I -T -w(\omega) -\Sigma_\text{Hxc}[G] -G^{-1}\right\|^2.
\end{eqnarray}
Note that multiple choices for the gradient norm are possible.
Moreover, one can also introduce the following definitions: 
\begin{eqnarray}
   \label{eq:grad_functional2}
   D[\rho] &=& \inf_{ \substack{G \rightarrow \rho \\ w} } D[w,G] \geq 0, \\[5pt]
   \nonumber
   \mathcal{A}_\rho &=& \big\{ (w,G) : G \rightarrow \rho, \,\,\, D[w,G]=D[\rho] \big\},
   \label{eq:grad_functional3}
\end{eqnarray}
where $D[\rho]$ is the infimum of the norm over the set of 
$G$'s providing a given occupied spectral density $\rho$, and over all possible external local and dynamical potentials $w$.
In order to make the search in Eq.~\eqref{eq:grad_functional2} consistent with the uniqueness of the potential when $\rho \in S_v$, $w$ is constrained to have a number of poles  
smaller than the number of Dyson orbitals in $\rho(\mathbf{r},\omega)$ (see Secs.~\ref{sec:proof3} and \ref{sec:mapping} for details).
Next, $\mathcal{A}_\rho$ is the set of $(w,G)$ pairs 
reaching the minimum of the norm $D[\rho]$ (i.e. the points where the infimum is reached by the norm functional, if any).
In the last definition, if the infimum is not reached 
the set $\mathcal{A}_\rho$ is empty.

Finally, within the above definitions, the SF energy functional in the constrained search (CS) formulation can be written as:
\begin{eqnarray}
   \label{eq:LLdef_SF}
             F^{\text{SF,CS}}[\rho] &=& \inf_{G \in \mathcal{A}_\rho} E^\text{K}[v_{\text{ext}}=0,G], 
   \\[5pt]
   \nonumber
   \widetilde{E}^{\text{SF}}[v_{\text{ext}},\rho] 
           \nonumber
            &=& \!\! \int d\omega d\mathbf{r}\, v_{\text{ext}} \rho + F^{\text{SF,CS}}[\rho], 
\end{eqnarray}
where $G \in \mathcal{A}_\rho$ refers to $G$'s from the pairs $(w,{G}) \in \mathcal{A}_\rho$. 
The $v_{\text{ext}}=0$ condition is just a way to disregard the energy term originating from the external potential.
When $\mathcal{A}_\rho$ is empty, one could for instance define the infimum to mean $+\infty$~\cite{note_infty}.  
In practice, given a spectral density $\rho$ we first search for the $(w,G)$ pairs such that $G \to \rho$ and
that grant the smallest norm for the gradient of the Klein functional $D[w,G]$, i.e., that are closest to physical solutions.  Finally, we evaluate
the term $E^\text{K}[ v_{\text{ext}}=0,G]$ for those $G$'s and take the infimum over the energy values.
 
Notably, the definitions of $E^{\text{SF}}$ in Eqs.~(\ref{eq:HKdef_SF},\ref{eq:HKdef_SF_var}) and~\eqref{eq:LLdef_SF} are identical
for occupied spectral densities in $S_v \subset S_G$ when both formulations are available. 
In fact, when a spectral density is $v$-representable it means there exists a pair $(v_\rho,G_\rho)$ which is a stationary
solution of the Klein functional. Moreover, the potential $v_\rho$ has to be unique because of the one-to-one correspondence
of Sec.~\ref{sec:one-to-one}. We also consider $G_\rho$ to be a unique solution (this has been recently questioned and extensively investigated~\cite{Kozik2015PhysRevLett,Eder2014arXiv,Stan2015NewJPhys,Rossi2015JPhysA:MathTheor,Schafer2016PhysRevB,Tarantino2017PhysRevB,Farid2021arXiv})
of the Dyson equation originating from the Klein functional (making it stationary).
The infimum over the norm of the gradient is then zero, $D[\rho]=0$, and $\mathcal{A}_\rho$ 
has just one element, the pair $(v_\rho,G_\rho)$. 
In view of the above discussion, the CS formulation represents a generalization
of Eqs.~(\ref{eq:HKdef_SF},\ref{eq:HKdef_SF_var}) also in the framework of the SF functional.


\section{Spectral non-interacting mapping}
\label{sec:KS-mapping}


Having established in the previous Section that we can write the total energy as a functional
of the occupied spectral function $\rho(\mathbf{r},\omega)$, we then want to obtain a variational scheme in the spirit of KS-DFT.
In order to do so we first introduce an auxiliary Green's function aimed at representing the occupied spectral density and then use it to rewrite the total energy functional into a KS-like form suitable for approximations and numerical solution of the problem. 

\subsection{Representation of the occupied spectral density}
%
In this Section we introduce a non-interacting auxiliary system, subject to a local and dynamical spectral potential (SP) $v^{\text{SP}}(\mathbf{r},\omega)$, aimed at representing the physical occupied spectral density, i.e. such that its Green's function $G^{\text{SP}}(\omega)$ carries the physical $\rho(\mathbf{r},\omega)$. In practice one has: 
\begin{eqnarray}
   \label{eq:GSP_def}
   G^{\text{SP}}(\mathbf{r},\mathbf{r}',\omega) &=& \left[ \omega I
   -T -v^{\text{SP}}(\omega)
                      \right]^{-1}_{\mathbf{r},\mathbf{r}'} \\
    &=& \sum_s \frac{f_s^{\text{SP}}(\mathbf{r})f_s^{\text{SP} *}(\mathbf{r}')}{\omega -\epsilon_s \mp i0^+}.
    \label{eq:GSP_def2}
\end{eqnarray}
In the second line, a spectral representation in terms of SP Dyson orbitals is also
provided, leading to a representation of the occupied spectral density as:
\begin{eqnarray}
  \rho(\mathbf{r},\omega) &=& \sum_s^{\text{occ}}  \left| f_s^{\text{SP}}(\mathbf{r})\right|^2 \, \delta(\omega-\epsilon_s). 
\end{eqnarray}
Notably, the above definitions parallel the introduction of the non-interacting ground-state (Slater determinant) of the KS system to represent the charge density in DFT. 

In terms of representability, according to Ref.~\cite{Ferretti2024PRB}, every well-behaved Green's function (i.e. having real poles and positive semi-definite residues) is "embedding non-interacting $v$-representable" \cite{Ferretti2024PRB}, i.e., there exists a non-interacting system with auxiliary degrees of freedom, that has the physical Green's function as a local (embedded) GF.
The embedding self-energy of such a system is in general dynamical and non-local, and therefore is not necessarily a $v^{\text{SP}}(\mathbf{r},\omega)$. Nevertheless, the local and dynamical SP potential described above can be obtained
from the solution of
a dynamical Sham-Schl\"uter equation (SSE)~\cite{Sham-Schluter1983PhysRevLett,Gatti2007PhysRevLett}.
A discussion of the dynamical SSE resulting from the present SP formulation is presented in App.~\ref{sec:app_dsse}.

While analytical and numerical solutions of dynamical SSEs  have been found for selected cases~\cite{Gatti2007PhysRevLett,Ferretti2014PhysRevB,Vanzini2018EPJB,Vanzini2019arXiv,Vanzini2020FaradayDiscuss}, a general criterion granting the existence of such solutions, i.e., an embedding non-interacting $v$-representability condition with a local dynamical potential, is not known to date. 
On a positive side, local and dynamical functionals such as Koopmans~\cite{Dabo2010PhysRevB,Ferretti2014PhysRevB, Linscott2023JChemTheoryComput} and dynamical Hubbard~\cite{Chiarotti2024PRR,Caserta2025arXiv} approaches seem generally to provide physical and meaningful solutions, while this is not the case, e.g., of second order reduced density matrix that has always been challenged by representability conditions~\cite{Mazziotti2012ChemRev,Mazziotti2012PhysRevLett}.

\subsection{Partitioning of the total energy functional}
%
Assuming an SP representation of the occupied spectral density as in Eq.~\eqref{eq:GSP_def}, one obtains the following chain of invertible maps:
\begin{equation}
  \label{eq:equiv_spectral_density}
  G_\rho(\omega)  \longleftrightarrow \rho(\mathbf{r},\omega) 
   \longleftrightarrow G^{\text{SP}}(\omega),
\end{equation}
where $G_\rho$, introduced in Sec.~\ref{sec:SF-universal-def}, is the physical interacting GF corresponding to the unique (local and dynamical) potential $v_\rho$ that has $\rho$ as occupied spectral density. 
In the following we exploit the SP representation in the total energy functional of Eq.~\eqref{eq:HKdef_SF_var}. 
In practice one wants to replace $G_\rho$ in the explicit expression of $F^{\text{SF}}$ with $G^{\text{SP}}$ as much as possible, treating the unknown remainder as a $xc$ functional to be approximated.

In particular, assuming to deal with spectral densities whose spectral potentials contain the poles of $v_\text{ext}$ (see below), one has:
\begin{eqnarray}
    \text{Tr}_\omega \left\{v_{\text{ext}}G_\rho\right\}
       = \text{Tr}_\omega \left\{v_{\text{ext}}G^{\text{SP}}\right\}
       = \int \!\! d\mathbf{r}d\omega \, v_{\text{ext}} \rho .
\end{eqnarray}
Moreover, the terms in the second line of the expression of the universal functional $F^{\text{SF}}$ in Eq.~\eqref{eq:HKdef_F_SF} can be rewritten as, 
\begin{eqnarray}
    F_{2,3}[G_\rho] &=& \text{Tr}_\omega \left\{I -G_0^{-1}G_\rho \right\} 
                        -\text{Tr}_\omega \left\{v_0 G_\rho\right\} 
          \nonumber  
            \\
          &=& \text{Tr}_\omega \left\{I -(\omega I -T)G_\rho \right\}, 
\end{eqnarray}
which results in
\begin{equation}
  F_{2,3}[G_\rho] = F_{2,3}[G^{\text{SP}}] + \text{Tr}_\omega \left\{T (G_\rho-G^{\text{SP}}) \right\}.
\end{equation}
Similarly, the TrLn term in the third line of Eq.~\eqref{eq:HKdef_F_SF}, $F_4,$ can be rearranged as:
\begin{eqnarray}
   F_4[G_\rho] &=& \text{Tr}_\omega \left\{\text{Ln}\, G_0^{-1}G_\rho\right\} \\ \nonumber
     &=& \text{Tr}_\omega \left\{\text{Ln}\, G_0^{-1}G^{\text{SP}}\right\} + \text{Tr}_\omega \left\{\text{Ln}\, G^{\text{SP},-1}G_\rho\right\}.
\end{eqnarray}

Given the expressions above, the total energy of Eqs.~\eqref{eq:HKdef_SF} and~\eqref{eq:HKdef_SF_var} can be rewritten as:
\begin{eqnarray}
     \label{eq:klein_SP}
     E^{\text{SF}}[v_{\text{ext}},G^{\text{SP}}] 
                        &=&
                        \text{Tr}_\omega \left\{v_{\text{ext}}G^{\text{SP}}\right\} + E^{\text{SP}}_{\text{Hxc}}[G^{\text{SP}}]        
     \\[5pt] \nonumber
                        &+& \text{Tr}_\omega \left\{I -G_0^{-1}G^{\text{SP}} \right\} 
                        -\text{Tr}_\omega \left\{v_0 G^{\text{SP}}\right\} \\[4pt]
                        &+& \text{Tr}_\omega \left\{\text{Ln}\, G_0^{-1}G^{\text{SP}}\right\} +
                        \text{Tr}_\omega \left\{h_0 G_0\right\},
        \nonumber
\end{eqnarray}
where we have defined:
\begin{eqnarray}
   \label{eq:ESP_HXC_def}
   E^{\text{SP}}_{\text{Hxc}}[G^{\text{SP}}] &=& \Phi_{\text{Hxc}}[G_\rho] + \text{Tr}_\omega \left\{T (G_\rho-G^{\text{SP}}) \right\}
   \nonumber \\[5pt]
   &+& \text{Tr}_\omega \left\{\text{Ln}\,  G^{\text{SP},-1} G_\rho\right\}.
\end{eqnarray}
In doing so every term that is still explicit in $G_\rho$ goes into the unknown $xc$-functional.
Moreover, in the equation above, $E^{\text{SP}}_{\text{Hxc}}$ depends on $\rho(\mathbf{r},\omega)$ via $G^{\text{SP}}$ in view of the equivalence maps in Eq.~\eqref{eq:equiv_spectral_density}.
The expression of $E^{\text{SP}}_{\text{Hxc}}$ accounts for the $\Phi$ functional evaluated using $G_\rho$, i.e., the physical interacting GF obtained from $\rho$ via the one-to-one mapping. Additionally, it also accounts for a corrections to the kinetic energy term (from $F_2$) when $G_\rho$ is replaced with 
$G^{\text{SP}}$, as well as for the variation of the $F_3$ term, $\text{Tr}_\omega \left\{\text{Ln}\, G_\rho\right\}$. 
As discussed in Sec.~\ref{sec:exactness_kinetic}, within the hypotheses of the \acronym{} construction, notably leading to real-valued Dyson orbitals, the kinetic energy correction is shown to be zero under mild conditions (while it would be needed in the presence of, e.g., spin-orbit coupling or magnetic fields).
Last, the correction to the $F_3$ term accounts for a difference in the occupied poles of $\Sigma(\omega,[G_\rho])$ and $v^{\text{SP}}(\omega)$, according to the analytical expression for the $\text{Tr}_\omega \text{Ln}[G_0^{-1} G]$ terms provided in Eqs.~(35) and (B11) of Ref.~\cite{Ferretti2024PRB}.

%

\subsection{Spectral self-consistent equations}
%
Once we have expressed the $E^{\text{SF}}$ functional using $G^{\text{SP}}$, Eq.~\eqref{eq:klein_SP}, we can exploit its stationarity to determine the occupied spectral density $\rho(\mathbf{r},\omega)$.
By the equivalence $G^{\text{SP}} \longleftrightarrow \rho(\mathbf{r},\omega)$, 
$E^{\text{SP}}_{\text{Hxc}}$ can be seen as a functional of $G^{\text{SP}}$ as well as of the spectral density. 
By introducing the definitions
\begin{eqnarray}
   \label{eq:vxc_SP_1}
   \frac{\delta E^{\text{SP}}_{\text{Hxc}}[\rho]}{\delta \rho(\mathbf{r},\omega)} &=& v^{\text{SP}}_{\text{Hxc}}(\mathbf{r},\omega), \\[5pt]
   \frac{\delta E^{\text{SP}}_{\text{Hxc}}[G^{\text{SP}}]}{\delta G^{\text{SP}}(\mathbf{r},\mathbf{r}',\omega)} &=& \frac{1}{2 \pi i}\Sigma^{\text{SP}}_\text{Hxc}(\mathbf{r}',\mathbf{r},\omega)
\end{eqnarray}
and
comparing the derivatives wrt $f_s^{\text{SP}*}(\mathbf{r})$ of the energy as a functional of $G^{\text{SP}}$ and of $\rho(\mathbf{r},\omega)$, we obtain:
\begin{eqnarray}
  & &\int \frac{\delta E^{\text{SP}}_{\text{Hxc}}[\rho]}{\delta \rho(\mathbf{r},\omega)} \, \frac{\delta \rho(\mathbf{r},\omega)}{\delta f_s^{\text{SP}*}(\mathbf{r})} d\omega  = v^{\text{SP}}_{\text{Hxc}}(\mathbf{r},\epsilon_s) f^{\text{SP}}_s(\mathbf{r}), \\[5pt]
  \nonumber
  & &\int \frac{\delta E^{\text{SP}}_{\text{Hxc}}[G^{\text{SP}}]}{\delta G^{\text{SP}}(\mathbf{r}_1,\mathbf{r},\omega)} \, 
  \frac{\delta G^{\text{SP}}(\mathbf{r}_1,\mathbf{r},\omega)}{\delta f_s^{\text{SP}*}(\mathbf{r})} d\omega d\mathbf{r}_1 = 
  \langle \mathbf{r} | \Sigma^{\text{SP}}_\text{Hxc}(\epsilon_s) | f_s \rangle.
\end{eqnarray}
Since the last two lines should be identical, 
we find that the SP Hxc self-energy at frequencies $\epsilon_s$ behaves as a local potential.

By taking into account the above discussion, the total energy in Eq.~\eqref{eq:klein_SP} is then made stationary by the GF:
\begin{equation}
   G^{\text{SP}} = \Big[\omega I -T -v_\text{ext}(\omega) -\Sigma^{\text{SP}}_\text{Hxc}(\omega,[G^{\text{SP}}]) \Big]^{-1},
\end{equation}
which, in turn, can be solved by solving self-consistently the following spectral equations:
\begin{eqnarray}
   \label{eq:dynamical_KS}
  \Big[ T +v_\text{ext}(\epsilon_s) +
        v^{\text{SP}}_{\text{Hxc}}(\epsilon_s) \Big] | f_s^{\text{SP}}\rangle  = \epsilon_s | f_s^{\text{SP}}\rangle, \qquad 
\end{eqnarray}
where the normalization of the Dyson SP orbitals is obtained as:
\begin{eqnarray}
   \langle f_s^{\text{SP}} |  f_s^{\text{SP}} \rangle = 1 + \langle f_s^{\text{SP}} | \dot{v}_{\text{ext}}(\epsilon_s) +  \dot{v}^{\text{SP}}_{\text{Hxc}}(\epsilon_s) | f_s^{\text{SP}} \rangle.
   \label{eq:SP_dyson_norm}
\end{eqnarray}
In the expressions above, $v_\text{ext}(\epsilon_s)$ and $v^{\text{SP}}_{\text{Hxc}}(\epsilon_s)$ are local potentials, and, comparing with Eq.~\eqref{eq:GSP_def}, we also have:
\begin{equation}
   \label{eq:vSP_decomposition} 
   v^{\text{SP}}(\mathbf{r},\epsilon_s) = v_\text{ext}(\mathbf{r},\epsilon_s) +
        v^{\text{SP}}_{\text{Hxc}}(\mathbf{r},\epsilon_s),
\end{equation}
and similarly for their frequency derivatives.
Equations~\eqref{eq:dynamical_KS} and~\eqref{eq:SP_dyson_norm} represent the third main result of this work and resemble the self-consistent Kohn-Sham 
equations 
on the one side, and the (self-consistent) Dyson equation of Green's function methods on the other.

At variance with the ordinary KS equations, here the SP Hxc potential, given by functional derivative of the three terms in the rhs of Eq.~\eqref{eq:ESP_HXC_def}, depends on the occupied spectral density, in turn described by the SP Dyson orbitals and Green's function according to Eqs.~\eqref{eq:GSP_def} and~\eqref{eq:GSP_def2}, highlighting the need to solve the problem self-consistently.
%
%
Moreover, these spectral self-consistent equations feature a dynamical potential
and take the form of a non-linear eigenvalue problem~\cite{Guttel2017ActaNumerica}, which we have recently addressed in a series of works~\cite{Chiarotti2022PRR,Chiarotti2023PhD,Ferretti2024PRB,Chiarotti2024PRR}. These latter introduce the algorithmic-inversion method on sum-over-poles, showing an effective and numerically precise way to deal with dynamical potentials: when these are expressed as sum-over-poles, an exact solution of the Dyson equation for the Green's function (also in the form of a sum-over-poles), can be obtained by solving a linear, rather than a non-linear, eigenvalue problem in a larger space, diagonalizing an appropriate extended Hamiltonian.

\subsection{Exactness of the $T^{\text{SP}}[\rho]$ kinetic energy functional}
\label{sec:exactness_kinetic}
%
In this Section we show that the kinetic energy correction appearing in Eq.~\eqref{eq:ESP_HXC_def} is zero for systems with real, non-degenerate (except for spin) Dyson orbitals, and discuss how this latter hypothesis can be removed. In the present construction, orbitals can be taken real in view of the Hamiltonian discussed in Sec.~\ref{sec:ham_embedding}, which is time-reversal symmetric and does not include spin-orbit coupling. 
In particular,  
the kinetic energy of the auxiliary SP system,
\begin{eqnarray}
  T^{\text{SP}}[\rho] &=& -\frac{1}{2} \sum_s^{\text{occ}} \int d\mathbf{r} \, f_s^{\text{SP} *}(\mathbf{r})\nabla^2 f_s^{\text{SP}}(\mathbf{r}), \\
  \nonumber
  &=& \text{Tr}_\omega \left\{ T G^{\text{SP}}(\omega) \right\},
\end{eqnarray}
is not in general granted to be the physical kinetic energy,  similarly to the case of KS-DFT. Remarkably, the following result holds:
\begin{theorem}
Given a spectral density that can be represented by means of an auxiliary SP system, the $T^{\text{SP}}[\rho]$ functional gives the exact kinetic energy provided that the Dyson orbitals in the physical and SP systems 
are real and non-degenerate (except for spin). 
\end{theorem}
{\it Proof.}
%
%
Since both the SP and physical Dyson orbitals are real valued by hypothesis, one can make use of the  
von Weizs\"acker kinetic energy functional~\cite{Weizsacker1935ZeitschriftfurPhysik,Acharya1980PNAS}, here applied to $\rho_s(\mathbf{r}) = |f_s(\mathbf{r})|^2$,
\begin{equation}
  T_{vW}[\rho_s] = \frac{1}{8} \int d\mathbf{r} \frac{ \nabla \rho_s(\mathbf{r}) \cdot \nabla \rho_s(\mathbf{r})}{ \rho_s(\mathbf{r})},
\end{equation}
which provides  
the exact kinetic energy of single-orbital densities when orbitals are real. Therefore, one can write:
\begin{equation}
   \langle f^{\text{SP}}_s | T | f^{\text{SP}}_s \rangle
      = T_{vW}[\rho_s] = \langle f_s | T | f_s \rangle,
\end{equation}
since by construction, the SP system has the same spectral density of the physical system. In a nutshell, the von Weizs\"acker functional protects the kinetic energy of the SP auxiliary system, proving the statement for non-degenerate orbitals.
$\Box$

In the presence of degenerate orbitals, one can  imagine a potential perturbation (even static and local) that breaks all possible spatial symmetries and the resulting degeneracies of the Dyson orbitals, while keeping the orbitals real.
Since the strength of the perturbation can be controlled and made infinitesimal while keeping the degeneracies broken, in any neighborhood of the spectral density $\rho(\mathbf{r},\omega)$ there are infinitely many points where the kinetic energy functional $T^{\text{SP}}[\rho]$ is exact. 
%
If one could assume that the SP potential and Dyson orbitals change continuously, then the kinetic energy functional $T^{\text{SP}}[\rho]$ would be exact also in the presence of degenerate orbitals.
In this regard, we expect that further numerical and theoretical analysis of the dynamical SSE in App.~\ref{sec:app_dsse} would be helpful to explore this point.
%
%
Moreover, while the above result is valid for the \acronym{} construction presented in this work, 
this is not the case, e.g., in the presence of spin-orbit coupling, where the von Weizs\"acker functional cannot be in general invoked even for non-degenerate orbitals, or magnetic fields.
Notably, the exactness of the SP kinetic energy functional somehow parallels the case of the (one-body) reduced-density matrix functional theory (RDMFT)~\cite{Gilbert1975PhysRevB,Lathiotakis2005PhysRevA}, where the kinetic energy is also explicitly known in terms of the reduced density matrix, and thereby exact by construction.
%

\subsection{Physical content of $E^{\text{SP}}_{\text{Hxc}}$ and approximations}
%
We are now in the position to discuss the physical content of the SP Hxc functional, which is the only term in Eq.~\eqref{eq:dynamical_KS} to be approximated.  
As usual, the Hartree term depends just on the total charge density $n(\mathbf{r})=\int_{-\infty}^\mu d\omega \,\rho(\mathbf{r},\omega) = \sum_s^{\text{occ}} \rho_s(\mathbf{r})$, and it can be separated out. 
Similar considerations are valid for the exchange term, which can be either treated locally (i.e., approximated by means of the occupied spectral density $\rho(\mathbf{r},\omega)$), or in an orbital-dependent fashion, exploiting the explicit knowledge of the SP Dyson orbitals $f^{\text{SP}}_s(\mathbf{r})$, and leading to a non-local potential.

The correlation term also needs to be approximated, but now as a functional of the occupied spectral density (a local but dynamical object) rather than the ground-state charge density (a local but static object) as was the case for KS-DFT. One immediate conceptual advantage that arises from a dynamical formulation is that the discontinuity with respect to the particle number $N$ of the exact exchange-correlation spectral functional is expected to become unnecessary~\cite{Ferretti2014PhysRevB}, relieving its approximations from 
the description of such a feature, necessary 
in the static exchange-correlation functional of exact KS-DFT~\cite{Perdew-Levy1983PRL,Tempel2009JChemTheoryComput,Helbig2009JChemPhys}.
At the same time, given that Dyson orbitals decay exponentially in vacuum (when accessible) according to the ionization potential $E_{IP} = E^{N-1}_0 -E^N_0 = -\epsilon_{s=0}$~\cite{Almbladh1985PhysRevB} (wrt the vacuum level, set to zero), while orbitals subject to a local potential as in Eq.~\eqref{eq:dynamical_KS} decay exponentially with their own eigenvalue $\epsilon_s$, we foresee the need for $v^{\text{SP}}(\mathbf{r},\epsilon_s)$ to possibly build a constant off-set in the vacuum region. Nevertheless, given the exponentially decaying densities, we expect that even working with approximations that do not reproduce the offset should not lead to drawbacks.

In terms of existing approaches and possible approximations, Koopmans' compliant functionals~\cite{Dabo2010PhysRevB,Linscott2023JChemTheoryComput} can be seen as providing an orbital-dependent, quasiparticle approximation to the SP Hxc potential~\cite{Ferretti2014PhysRevB}. 
Interestingly, in a series of recent works~\cite{Vanzini2018EPJB,Vanzini2019arXiv,Vanzini2020FaradayDiscuss,Aouina2020FaradayDiscuss}, Vanzini, Gatti, Reining and coworkers have proposed a general framework (connector theory) to devise functional approximations targeting selected observables as based on existing physical knowledge, and have explicitly considered the case of photoemission observables and local dynamical potentials.
Moreover, fully dynamical and local potentials arise in the model self-energies of Vanzini and Marzari~\cite{Vanzini2023arXiv} and the dynamical Hubbard functional of Chiarotti {\it et al.}~\cite{Chiarotti2023PhD,Chiarotti2024PRR}. These formulations, despite their simplicity, lead to a remarkable accuracy in the evaluation of thermodynamic and spectroscopic properties~\cite{Colonna2019JChemTheoryComput,Nguyen2018PhysRevX,Vanzini2023arXiv,Chiarotti2024PRR}. 
In passing, we note that all these formulations use screening as physical input (in the form of orbital-dependent screening coefficient $\alpha_i$ in Koopmans, or $U(\omega)$ in Refs.~\cite{Chiarotti2023PhD,Chiarotti2024PRR}). 

Last, the extension to spin-dependent quantities or relativistic corrections --- not discussed here ---
can be envisioned along the lines of the original introduction of spin-resolved functionals and spin-orbit coupling~\cite{Jones1989RMP}. Interestingly, Koopmans' compliant functionals were recently extended to the non-collinear relativistic case~\cite{Marrazzo2024PRR}, showing again very good agreement with experiments.


\begin{table*}[t]
\centering
\renewcommand{\arraystretch}{1.25}
\begin{tabular}{p{.42\linewidth} p{.45\linewidth} p{.10\linewidth}}
\hline
\textbf{DFT} & \textbf{Spectral-DFT} & \\
\hline
\\[-6pt]
\multicolumn{2}{c}{\textbf{Fundamental variable}} \\[0.3em]
Ground-state charge density $n(\mathbf r)$ &
Occupied spectral density $\rho(\mathbf r,\omega)$ \\[4pt]
$n(\mathbf{r})=\sum_{i} |\phi_i(\mathbf r)|^2$ & 
$\rho(\mathbf{r},\omega)=
\sum_{s} |f_s(\mathbf r)|^2\,\delta(\omega-\epsilon_s)$
&
Eq.~\eqref{eq:spectral_density}
\\[0.7em]
%
%
\multicolumn{2}{c}{\textbf{One-to-one correspondence}} \\[0.3em]
$n(\mathbf r)\;\longleftrightarrow\;v_{\mathrm{ext}}(\mathbf r)$ &
$\rho(\mathbf r,\omega)\;\longleftrightarrow\;v_{\mathrm{ext}}(\mathbf r,\omega)$ 
& Fig.~\ref{fig:potential_density_mapping}
\\[0.7em]
%
%
\multicolumn{2}{c}{\textbf{Variational functional}} \\[0.3em]
$E^{\text{DFT}}[n]=\int v_{\mathrm{ext}}\,n+F^{\text{DFT}}[n]$ &
$E^{\mathrm{SF}}[\rho]=\iint v_{\mathrm{ext}}\,\rho+F^{\mathrm{SF}}[\rho]$ 
& Eqs.~\eqref{eq:FT_partition},~\eqref{eq:HKdef_SF_var}
\\[0.7em]
%
%
\multicolumn{2}{c}{\textbf{Self-consistent one-body equations}} \\[0.3em]
$\Big[-\tfrac{1}{2}\nabla^2+v_{\mathrm{ext}}(\mathbf{r})+v^{\text{KS}}_{\text{Hxc}}(\mathbf{r})\Big]\phi_i(\mathbf{r})=\epsilon_i\phi_i(\mathbf{r})$ 
&
$\Big[-\tfrac{1}{2}\nabla^2+v_{\mathrm{ext}}(\mathbf{r},\epsilon_s)+
v^{\mathrm{SP}}_{\text{Hxc}}(\mathbf{r},\epsilon_s)\Big] f^{\mathrm{SP}}_s(\mathbf{r})
= \epsilon_s f^{\mathrm{SP}}_s(\mathbf{r}),$ 
& 
Eq.~\eqref{eq:dynamical_KS}
\\[6pt]
$\langle \phi_i|\phi_j\rangle=\delta_{ij}$
&
$\langle f^{\mathrm{SP}}_s|f^{\mathrm{SP}}_s\rangle
=1+\langle f^{\mathrm{SP}}_s|\dot v_{\mathrm{ext}}(\epsilon_s)
+\dot v^{\mathrm{SP}}_{\text{Hxc}}(\epsilon_s)|f^{\mathrm{SP}}_s\rangle$
&
Eq.~\eqref{eq:SP_dyson_norm}
\\[6pt]
$v^{\text{KS}}_{\text{Hxc}}(\mathbf{r};[n]) = {\delta  E^{\text{KS}}_{\text{Hxc}}[n] }/ {\delta n(\mathbf{r})}$
&
$v^{\text{SP}}_{\text{Hxc}}(\mathbf{r},\omega;[\rho]) = {\delta  E^{\text{SP}}_{\text{Hxc}}[\rho] }/ {\delta \rho(\mathbf{r},\omega)}$
& 
Eq.~\eqref{eq:vxc_SP_1}
\\[0.8em]
%
%
\hline
\end{tabular}
\caption{
A comparison between DFT~\cite{Hohenberg-Kohn1964PR, Kohn-Sham1965PR, Eschrig2003book} and \acronym{} (this work). 
In the DFT column, the index $i$ runs over Kohn-Sham orbitals, while in the \acronym{} column, the $s$ indices span the occupied Dyson orbitals, with $\epsilon_s$ describing the physical charged excitations.
The one-to-one mapping of \acronym{} is more general than DFT, allowing for local and dynamical external potentials
$v_{\mathrm{ext}}(\mathbf r,\omega)$.
At variance with the KS construction, in \acronym{} a local and dynamical potential ($v_{\mathrm{ext}}+
v^{\mathrm{SP}}_{\text{Hxc}}$) is used to represent the spectral density, leading to dynamical self-consistent equations. Given the faithful representation of the exact spectral density, the kinetic energy is also exact in the SP system under mild assumptions (See Sec.~\ref{sec:exactness_kinetic} for a in-depth discussion).
\label{tab:rosetta}
}
\end{table*}


\section{Conclusions}
\label{sec:conclusions}
In this work we have formulated a functional theory --- \acronym{} --- of the (occupied) spectral density $\rho(\mathbf{r},\omega)$ rather than the total charge density $n(\mathbf{r})$, as done in standard density-functional theory. The \acronym{} formulation provides simultaneously total energies and charged excitations, and can therefore be naturally applied to interpret relevant spectroscopies, as already envisioned by Sham and Kohn in Ref.~[\onlinecite{Sham1966PR}], and is free of any quasiparticle approximation. It can be seen as the next step in a three-pronged hierarchy (see Box~1 of Ref.~[\onlinecite{Marzari2021NatMater}] for a graphical representation) where TDDFT (that addresses neutral excitations) is the generalization of DFT with respect to time, \acronym{} is the generalization of DFT with respect to frequency, and RDMFT is the generalization of DFT with respect to many-body wavefunction contractions. In doing so, while being oriented to describe correlations and excitations, we believe it strikes an optimal balance of providing a much needed frequency dependence in the functionals (and thus an electronic-structure theory that refers to real electrons) while preserving locality. This latter provides physical transparency~\cite{Kohn1996PRL} and likely avoids the challenges of representability that other exact functionals (even explicit ones, such as that of second-order reduced density matrix~\cite{Coleman1963RMP,Mazziotti2012ChemRev}) face.
In addition, we note that local and dynamical spectral potentials are not expected to feature strong derivative discontinuity features, as stemming, e.g., from early numerical results~\cite{Ferretti2014PhysRevB}.

A general formulation of the problem is provided, starting from the definition of the local and dynamical external potential corresponding to the occupied spectral density, interpreted as an embedding self-energy.
First, we provide in Sec.~\ref{sec:map_invertibility} a one-to-one correspondence between an occupied spectral density and its local dynamical external potential. Then, a definition of a universal functional of the spectral density is given in Eqs.~\eqref{eq:HKdef_SF} and~\eqref{eq:HKdef_SF_var}. Finally, spectral self-consistent equations are introduced in Eqs.~\eqref{eq:dynamical_KS} and~\eqref{eq:SP_dyson_norm}. 
We provide in Tab.~\ref{tab:rosetta} a summary of the parallels and differences of DFT and \acronym. 

In conclusion, we have generalized the theorems of DFT and the KS construction and equations to the next natural step, with a dependence on the occupied spectral density (local and dynamical) rather than the charge density (local and static). This leads to a dynamical formulation that enables to capture accurately and inexpensively spectral properties, in turn allowing for more physical and accurate expressions of the total energy. Being grounded in a theory of real electronic states, it removes some numerical and conceptual challenges --- such as the derivative discontinuity as a function of the number of electrons  --- of exact KS-DFT. 

It is hoped that this formulation will prompt further theoretical and numerical studies, and the development and application of ever more refined approximate spectral functionals~\cite{Chiarotti2023PhD,Chiarotti2024PRR}, complementing the formal efforts and great progress made by many-body perturbation theory~\cite{Onida2002RevModPhys} and dynamical mean-field theory~\cite{Georges1996RevModPhys,Kotliar2006RevModPhys}, to understand, predict, or design fundamental and applied properties in condensed-matter physics and in all the fields that have greatly benefited up to now from the power, accuracy, and computational ease of density-functional theory~\cite{vanNoorden2014Nature,Marzari2021NatMater}.


\begin{acknowledgments}
We thank Tommaso Chiarotti for a careful reading of the manuscript and for many illuminating scientific discussions on the subject. We also acknowledge all important discussions with L. Reining, M. Gatti, M. Vanzini, S. Baroni, D. Varsano, S. Pittalis, M. Quinzi, and M. Rontani.
We acknowledge financial support from 
the Swiss National Centre of Competence in Research MARVEL for ``Computational Design and Discovery of Novel Materials''.
\end{acknowledgments}

\appendix
%


\section{Location of Green's function zeros}
\label{sec:math_details}
%
In this work we make large use of the assumption of having the poles of $G$ (and of any other time-ordered
frequency dependent operator such as $\Sigma_{\text{Hxc}}$ or $v_{\text{ext}}$) to be {\it discrete}. 
This can be formally achieved by fitting the physical system of interest into a finite box of arbitrary size
(eg with periodic boundary conditions), meaning that a three-dimensional torus $\mathbb{T}^3$ 
is used as coordinate space instead of $\mathbb{R}^3$, as described e.g. in Ref.~\cite{Eschrig2003book}.

%
%
When performing integrals alike 
the first line of Eq.~(\ref{eq:vext_expectation}), i.e.
\begin{equation}
  \Delta E = \int \frac{d\omega}{2 \pi i} \text{Tr} \left\{ \Sigma(\omega) G(\omega) \right\},
\end{equation}
using the theorem of residues, poles from both the GF and the self-energy should be considered.
Nevertheless, if $G$ is the solution of a Dyson equation involving $\Sigma$,
the contributions from the (discrete) poles of $\Sigma$ can be neglected. This can be understood e.g.
by writing the Dyson equation as
\begin{equation}
\label{eq:dyson_reversed}
G_0^{-1}G - I = \Sigma G.
\end{equation}
Since the poles on the {\it rhs} must be the same as those on the {\it lhs}, the poles of $\Sigma$
must be cancelled by zeros of $G$.
This result, often invoked in the literature, can be summarized by the following statement.

\begin{lemma}
\label{theo:polesG}
Given $\Sigma$, $G_0$ and $G$, expressed as sums of isolated poles, and connected by a
Dyson equation as in Eq. \eqref{eq:dyson_reversed}, then the poles of $\Sigma$ are matched by zero's of $G$.
\end{lemma}

\noindent {\it Proof.}
According to the hypothesis, the self-energy can be written as
\begin{equation}
   \Sigma(\mathbf{x},\mathbf{x}',\omega) = \Sigma_0(\mathbf{x},\mathbf{x}') 
          + \sum_s \frac{\Gamma_s(\mathbf{x},\mathbf{x}')}{\omega -\Omega_s}, 
\end{equation}
where $\Omega_s$ may be taken complex.
Via the Dyson equation, we want to evaluate $G(\omega\to\Omega_{j})$:
\begin{eqnarray}
    \label{eq:dyson_theo}
    G(\omega) &=& \left[ C(\omega) -\frac{\Gamma_j}{\omega-\Omega_j} 
    \right]^{-1}, \\[5pt]
         C(\omega) &=&  \omega I -h_0 -\Delta\Sigma(\omega),
    \label{eq:dyson_theo_C}
\end{eqnarray}
$\Delta\Sigma(\omega)$ being the total self-energy except for the $\Omega_j$ contribution:
\begin{equation}
  \Delta\Sigma(\omega) = \Sigma(\omega) - \frac{\Gamma_j}{\omega-\Omega_j}.
\end{equation}
If $\Gamma_j$  
is invertible, the limit $\omega \to \Omega_j$ is trivially 
taken resulting in 
\begin{equation}
G(\omega) \sim \,  -(\omega-\Omega_j)\, \Gamma^{-1}_j \, \to 0 ,
\end{equation}
which is the thesis.
Moreover, one also has:
\begin{equation}
   G(\omega) (\omega-\Omega_j)^{-1} \to -\Gamma_j^{-1} \quad \text{for \ } \omega \to \Omega_j .
   \label{eq:G_times_omega_minus_Omega}
\end{equation}
Loosely speaking, any pole of $\Sigma$ would appear in the denominator of the GF, thereby resulting in a
zero of $G$. 

Instead, if $\Gamma_j$ has a null subspace, we can
introduce the $Q_j$ projector associated with it, such that $I=Q_j+P_j$. On the $Q_j$ subspace the
effect of the $\Omega_j$ pole is cancelled, so care must be used.
The inversion operation of Eq.~(\ref{eq:dyson_theo}) can then be performed block-wise over the 2$\times$2 block matrix
defined by the $P_j$ and $Q_j$ projectors, following the same algebra used 
e.g. to compute the embedding self-energy of a system coupled to a bath~\cite{Martin-Reining-Ceperley2016book,Ferretti2024PRB}.
Considering the definition of $C(\omega)$ from Eq.~\eqref{eq:dyson_theo_C}, one can write:
\begin{eqnarray} 
   \Bigg(
   \begin{matrix} 
      C_{PP} -\frac{\Gamma_j}{\omega-\Omega_j} &  C_{PQ} \\
      C_{QP} & C_{QQ}
   \end{matrix}
   \Bigg)
   \Bigg(
   \begin{matrix} 
      G_{PP} &  G_{PQ} \\
      G_{QP} &  G_{QQ}
   \end{matrix}
   \Bigg)
   =
   I,
\end{eqnarray}
leading to the following equations:
\begin{eqnarray}
   & &\left[ C_{PP} -\frac{\Gamma_j}{\omega-\Omega_j} -C_{PQ} C^{-1}_{QQ} C_{QP} \right] G_{PP} = I_P 
   \nonumber \\[4pt] 
   & &G_{QP} = - C^{-1}_{QQ} C_{QP} G_{PP}
   \nonumber \\ 
   & &G_{PQ} = -\left[ C_{PP} -\frac{\Gamma_j}{\omega-\Omega_j}\right]^{-1} C_{PQ} G_{QQ}
   \nonumber \\
   & &\left[ C_{QQ} -C_{QP} \left[ C_{PP}-\frac{\Gamma_j}{\omega-\Omega_j}\right]^{-1} C_{PQ}\right] G_{QQ} = I_Q .
   \nonumber
\end{eqnarray}
In the $\omega \to \Omega_j$ limit, one has:
\begin{equation}
   \left[ C_{PP} -\frac{\Gamma_j}{\omega-\Omega_j}\right]^{-1} \sim -(\omega-\Omega_j) \Gamma_j^{-1},
\end{equation}
which gives
\begin{eqnarray}
   G_{PP}(\omega) &\sim& -(\omega-\Omega_j) \Gamma_j^{-1} \to 0 
   \\
   G_{QP} &\to& 0
   \nonumber \\
   G_{PQ} &\to& 0
   \nonumber \\
   G_{QQ} &\to& C_{QQ}^{-1}(\Omega_j).
   \nonumber
\end{eqnarray}
In other words, by taking the $\omega \to \Omega_j$ limit one finds that
$G$ is block diagonal on $P_j$ and $Q_j$ and that $G_{PP}\to 0$ while $G_{QQ}(\omega)$ stays finite.
This finally proves that the poles of $\Sigma$ are matched by zeros of $G$, at least out of the null subspace of each residue $\Gamma_j$.
$\Box$

As an important consequence, 
integrals of the form $\text{Tr}_\omega \left\{ \Sigma G\right\}$, e.g. as those in Eq.~\eqref{eq:vext_expectation}, do not have any explicit contribution from the poles of $\Sigma$. In fact:
\begin{eqnarray}
   I &=& \int \frac{d\omega}{2 \pi i} e^{i \omega 0^+} \, \text{Tr} \big\{ \Sigma G \big\} 
     \\[4pt] \nonumber
     &=& \sum_s^{\text{occ}} \text{Tr} \big\{ \Sigma(\epsilon_s) A_s \big\}+ 
         \sum_s^{\text{occ}} \text{Tr} \big\{\Gamma_j G(\Omega_j) \big\}
     \\ \nonumber
     &=& \sum_s^{\text{occ}} \text{Tr} \big\{\Sigma(\epsilon_s) A_s \big\}.
\end{eqnarray}
Moreover, 
at any frequency where the spectral density is non-zero (i.e., where there
are poles of the GF associated to Dyson orbitals), the corresponding dynamical potentials 
(both the external potential and the interaction self-energies) cannot have any poles, unless their residues spanning a subspace overlapping with the Dyson orbitals. In other words, $\Sigma(\epsilon_s)$ must be Hermitian.
In fact, assuming to have a Dyson orbital with energy $\epsilon_s = \Omega_j$ and not in the null subspace of the corresponding residue $\Gamma_j$ of the self-energy, one would obtain $G(\epsilon_s)=0$ on $P_j$, contradicting the hypothesis of having a non-zero spectral density.


\section{Uniqueness of the rational interpolation of the potential}
\label{sec:math_uniqueness_rational}
%
Let us consider a local dynamical potential expressed as a sum-over-poles, i.e., as a rational function, of the form:
\begin{eqnarray}
    \label{eq:v_rational}
    v(\mathbf{r},z) &=& v_0(\mathbf{r}) + \sum_n^N \frac{R_n(\mathbf{r})}{z-\Omega_n}  
    \\ \nonumber
    &=& \frac{\sum_{p=0}^N a_p(\mathbf{r})z^p} {\sum_{q=0}^N b_q z^q} = \frac{P^N_{\mathbf{a}(\mathbf{r})}(z)}{P^N_\mathbf{b}(z)},
\end{eqnarray}
where $P^N_{\mathbf{a}}(z)$ and $P^N_{\mathbf{b}}(z)$ are polynomials of degree $N$ in $z$ with the $N+1$ coefficients $a_0,\dots,a_N$ (also depending on $\mathbf{r}$) and $b_0,\dots,b_N$, respectively. Without loss of generality we can set $b_N=1$.
Since the ${a}_p(\mathbf{r})$ coefficients depend on $\mathbf{r}$, the above expression is a vector-valued rational function.
The frequency derivative of the potential can then be written as:
\begin{equation}
   \label{eq:v_interpolation_deriv}
  \dot{v}(\mathbf{r},z)  = \frac{\partial \,{v}(\mathbf{r},z)}{\partial z} = \frac{\dot{P}^N_{\mathbf{a}(\mathbf{r})}(z)}{P^N_{\mathbf{b}}(z)} -v(\mathbf{r},z) \frac{\dot{P}^N_{\mathbf{b}}(z)}{P^N_{\mathbf{b}}(z)}.
\end{equation}

According to the analysis presented in Sec.~\ref{sec:mapping},
in this Appendix we discuss the uniqueness of the potential once it is known in a set of points $z_i$, possibly together with its averaged first derivative at the same points (which falls within the class of vector-valued rational interpolation problems~\cite{Graves-Morris1983NumerischeMathematik}).
In detail, as also defined in Eq.~\eqref{eq:dynamical_sampling}, we set:
\begin{eqnarray}
 \label{eq:v_interpolation_sampling}
 v_i(\mathbf{r}) &=& v(\mathbf{r},z_i),
 \\ \nonumber
 \dot{\overline{v}}_i &=& \int d\mathbf{r} \, \dot{v}(\mathbf{r},z_i) \,w_i(\mathbf{r}),
\end{eqnarray}
where  
$w_i(\mathbf{r})$ is an averaging function (as stressed by the bar in $\dot{\overline{v}}_i$).
This means that we need to prove the uniqueness of the interpolation using the rational form of Eq.~\eqref{eq:v_rational} given the sampling points in Eqs.~\eqref{eq:v_interpolation_sampling}.
In the following we first discuss the case of scalar potentials (where the dependence on $\mathbf{r}$ is dropped), and then extend the results to the space-dependent (vector-valued) case of Eq.~\eqref{eq:v_rational}.

\subsection{Rational interpolation: scalar case}
%
Given $N$ the number of poles (or equivalently the degree of the $P_\mathbf{a}^N$ and $P_\mathbf{b}^N$ polynomials) in Eq.~\eqref{eq:v_rational}, one can uniquely determine the potential by providing $2N+1$ values $v_i$ (i.e. sampling only the potential). This is known as Cauchy or rational interpolation problem~\cite{Macon1962TheAmericanMathematicalMonthly,
Antoulas1988LinearAlgebraanditsApplications,Trefethen2019book_ch26}.
The existence of an interpolating rational function is not granted in general (see Ref.~\cite{Cortadella2018LinearAlgebraanditsApplications} for a recent description of the set of problematic cases), but if the solution exists it is unique.
This can be proven, e.g., following Ref.~\cite{Macon1962TheAmericanMathematicalMonthly}.

Let us assume two rational functions with the desired property exist, namely
\begin{equation}
   \label{eq:v_interpolation_vi_12}
   v_i = \frac{P^N_{\mathbf{a}_1}(z_i)}{P^N_{\mathbf{b}_1}(z_i)} = \frac{P^N_{\mathbf{a}_2}(z_i)}{P^N_{\mathbf{b}_2}(z_i)},  \qquad \quad \forall z_i
\end{equation}
where $P^N_{\mathbf{a}_i}(z)$ and $P^N_{\mathbf{b}_i}(z)$ are taken to be relatively prime (no common roots) for $i=1,2$.
This means that the polynomial 
\begin{eqnarray}
   \label{eq:T_poly}
   T(z) = P^N_{\mathbf{a}_1}(z) P^N_{\mathbf{b}_2}(z) -P^N_{\mathbf{a}_2}(z) P^N_{\mathbf{b}_1}(z),
\end{eqnarray}
which is of degree at most $2N$, has at least $2N+1$ roots and is therefore identically zero. In turn this also implies that
\begin{equation}
P^N_{\mathbf{a}_1}(z) = \frac{P^N_{\mathbf{a}_2}(z) P^N_{\mathbf{b}_1}(z)} { P^N_{\mathbf{b}_2}(z)},
\end{equation}
meaning that $P^N_{\mathbf{b}_2}(z)$, being relatively prime to $P^N_{\mathbf{a}_2}(z)$, is a divisor of $P^N_{\mathbf{b}_1}(z)$. The viceversa is also true ($P^N_{\mathbf{b}_1}$ divides $P^N_{\mathbf{b}_2}$), implying that $P^N_{\mathbf{b}_1}(z) = P^N_{\mathbf{b}_2}(z)$ and $P^N_{\mathbf{a}_1}(z) = P^N_{\mathbf{a}_2}(z)$.

\subsection{Rational Hermite interpolation: scalar case}
%
As variance with the previous Section, the interpolation problem is now generalized by providing not just the values $v_i$ of the potential at $z_i$, but also the values of its first derivative $\dot{v}_i$.
For instance, the problem is determined (actually over-determined) by providing $v_i$ and $\dot{v}_i$ at $N+1$ points, and undergoes the name of rational Hermite interpolation.
Also in this case, the existence of the solution is not granted, but when it exists, it is unique.

Uniqueness can be proven as follows.
Let us assume there are two rational functions fulfilling the conditions on $v_i$ and $\dot{v}_i$. Equation~\eqref{eq:v_interpolation_vi_12} is still valid (though on $N+1$ points $z_i$). Moreover, equating the expression for the derivative of $v(z)$ from Eq.~\eqref{eq:v_interpolation_deriv} (disregarding the $\mathbf{r}$ index), we have:
\begin{eqnarray}
   \nonumber
   \frac{\dot{P}^N_{\mathbf{a}_1}(z_i)}{P^N_{\mathbf{b}_1}(z_i)} -v(z_i) \frac{\dot{P}^N_{\mathbf{b}_1}(z_i)}{P^N_{\mathbf{b}_1}(z_i)} = 
   \frac{\dot{P}^N_{\mathbf{a}_2}(z_i)}{P^N_{\mathbf{b}_2}(z_i)} -v(z_i) \frac{\dot{P}^N_{\mathbf{b}_2}(z_i)}{P^N_{\mathbf{b}_2}(z_i)}, 
\end{eqnarray}
which, after some rearrangements and making use of Eq.~\eqref{eq:v_interpolation_vi_12}, becomes
\begin{multline}
   \label{eq:T_poly_deriv}
   \dot{P}^N_{\mathbf{a}_1}(z_i) P^N_{\mathbf{b}_2}(z_i) -P^N_{\mathbf{a}_2}(z_i) \dot{P}^N_{\mathbf{b}_1}(z_i) 
      \\[4pt]
  - \dot{P}^N_{\mathbf{a}_2}(z_i) P^N_{\mathbf{b}_1}(z_i) + P^N_{\mathbf{a}_1}(z_i) \dot{P}^N_{\mathbf{b}_2}(z_i) = 0.
\end{multline}
The above expression means that the derivative $\dot{T}(z)$ of the polynomial in Eq.~\eqref{eq:T_poly} is zero at each $z_i$, together with the values of the polynomial itself, $T(z_i)=0$.
Therefore, $T(z)$ (at most of degree $2N$) has $N+1$ double roots, making $T(z)$ identically null. The proof is then completed as in the previous case.

\subsection{Generalization to space-dependent potentials}
\label{sec:math_uniqueness_rational_space_dep}
%
The extension of the uniqueness results presented above to the case of space-dependent potentials $v(\mathbf{r},z)$ is straightforward when the sampling is performed only on the values of the potential: with $2N+1$ sampling points $v_i(\mathbf{r})$, the uniqueness of the rational interpolation is granted for potentials where both $P_\mathbf{a}(z)$ and $P_\mathbf{b}(z)$ depend parametrically on $\mathbf{r}$, that is a case even more general than the potentials in Eq.~\eqref{eq:v_rational}.

When discussing Hermite-like rational interpolation, according to Eq.~\eqref{eq:v_interpolation_sampling}, in the physical formulation of interest here, the samples of the potential derivative are averaged by the function $w_i(\mathbf{r})$.
Assuming two solutions exists, we obtain a spatial dependent version of Eq.~\eqref{eq:v_interpolation_vi_12}.
By defining the averaged quantities:
\begin{eqnarray}
   \overline{v}_i &=& \int d\mathbf{r} \, v(\mathbf{r},z_i) \,w_i(\mathbf{r}), 
   \nonumber \\ 
   {P}^N_{\overline{\mathbf{a}}}(z_i) &=& \int d\mathbf{r} \, {P}^N_{\mathbf{a}(\mathbf{r})}(z_i) \, w_i(\mathbf{r}),
\end{eqnarray}
we also obtain the averaged (scalar) version of Eq.~\eqref{eq:v_interpolation_vi_12} and Eq.~\eqref{eq:T_poly_deriv}. As for the scalar case, this implies that $P^N_{\mathbf{b}_1}(z) = P^N_{\mathbf{b}_2}(z)$ and, by means of Eq.~\eqref{eq:v_interpolation_vi_12} with space dependence, also $P^N_{\mathbf{a}_1(\mathbf{r})}(z_i) = P^N_{\mathbf{a}_2(\mathbf{r})}(z_i)$ at $N+1$ points $z_i$. Since both polynomials are at most of degree $N$, this implies the thesis and concludes the proof.


\section{Perturbation matrix elements in the embedding framework}
\label{sec:app_matrix_elements}
%
With reference to the embedding construction presented in Eqs.~\eqref{eq:full_hamiltonian}, one can explicitly evaluate selected matrix elements of $\delta \bar{V}_{\text{ext}}$, obtained by allowing for the variation of $\delta v_0(\mathbf{r}), \delta v_{0n}(\mathbf{r}), \delta \Omega_n$, with $n>0$. This is especially useful in the discussion of the potential-density map invertibility.

As a first step we introduce the generalized amplitudes:
\begin{eqnarray}
   F^{N}_{sm}(\mathbf{r}n) =
   \langle N-1,s | \hat{\psi}(\mathbf{r}n) | N,m\rangle\,
\end{eqnarray}
which reduce to the definition of Dyson orbitals when $m=0$ and $|N,0\rangle$ corresponds to the $N$-particle ground state of the system, i.e. $f_s(\mathbf{r}n)=F^N_{s0}(\mathbf{r}n)$. 
These quantities have been defined making reference to the $\hat{\psi}(\mathbf{r}n)$ operators with $n \geq 0$, but an equation of motion similar to Eq.~\eqref{eq:EOM_bath} can be derived,
\begin{equation}
    \label{eq:EOM_bath_2}
    \big[ E^{N}_m - E^{N-1}_s -\Omega_n\big] 
    F^{N}_{sm}(\mathbf{r}n) = v_{n0}(\mathbf{r}) F^{N}_{sm}(\mathbf{r}0), 
\end{equation}
which allows one to relate the values of $F^{N}_{sm}(\mathbf{r}n)$ in the auxiliary space ($n\ge1$) to those in the physical space $F^{N}_{sm}(\mathbf{r}0)$, at least when the coupling $v_{n0}(\mathbf{r})$ is non-zero.

Using Eqs.~\eqref{eq:full_hamiltonian} together with Eq.~\eqref{eq:EOM_bath_2}, by direct evaluation one obtains:
\begin{widetext}
\begin{eqnarray}
   \label{eq:Nm_dvext_N0}
   \langle N,m | \delta \bar{V}_{\text{ext}}| N,0\rangle &=&
   \sum_{s} \int d\mathbf{r} \,
   F^{N*}_{sm}(\mathbf{r}) \, 
   f_{s}(\mathbf{r}) \,
   \delta W^N_{m0,s}(\mathbf{r}),
   \\
   \langle N-1,s | \delta \bar{V}_{\text{ext}}| N-1,p\rangle &=&
   \sum_{t} \int  d\mathbf{r} \,
   F^{N-1,*}_{ts}(\mathbf{r}) \, 
   F^{N-1}_{tp}(\mathbf{r}) \,
   \delta W^{N-1}_{sp,t}(\mathbf{r}),
\end{eqnarray}
where we have defined:
%
\begin{eqnarray}
   \delta W^N_{mm',s}(\mathbf{r}) &=& \delta v_0(\mathbf{r}) + 
      \sum_n \Bigg[ \frac{\delta v_{n0}(\mathbf{r}) v_{0n}(\mathbf{r})}{\Delta E^N_{m's} -\Omega_n} 
      +\frac{v_{n0}(\mathbf{r})\delta v_{0n}(\mathbf{r})}{\Delta E^N_{ms} -\Omega_n} 
      + \frac{R_n(\mathbf{r}) \delta \Omega_n}{(\Delta E^N_{m's}-\Omega_n)(\Delta E^N_{ms} -\Omega_n)}
      \Bigg],
\end{eqnarray}
\end{widetext}
with $\Delta E^{N}_{ms} = E^N_m -E^{N-1}_s$, which gives $\epsilon_s = \Delta E^N_{0s}$.
Notably, when evaluating diagonal matrix elements, one can exploit the definition of $\delta v_{\text{ext}}(\mathbf{r},\omega)$ from Eq.~\eqref{eq:vext_variation}, and write
\begin{eqnarray}
  \langle N,0 | \delta \bar{V}_{\text{ext}}| N,0\rangle =
   \sum_{s} \int d\mathbf{r} \,
   \rho_{s}(\mathbf{r}) \,
   \delta v_{\text{ext}}(\mathbf{r},\epsilon_s) .
\end{eqnarray}


\section{Regular and null Dyson orbital contributions to $\delta v_{\text{ext}}$}
\label{sec:regular_dyson_orbitals_dvext}
%
As mentioned in Sec.~\ref{sec:regular_null_dyson}, Dyson orbitals can be categorized as regular or null, depending on whether they have finite norm or are identically zero, respectively. The latter case may arise, e.g., in the presence spin or other symmetries of the problem. 
Considering a generic single particle basis $\{\phi_m(\mathbf{r}n)\}$ in the extended space of Fig.~\ref{fig:hamiltonian} and the related annihilation operators
\begin{equation}
    \hat{c}_n = \sum_n \int d\mathbf{x} \,\phi^*_m(\mathbf{x}n) \, \hat{\psi}(\mathbf{x}n), 
\end{equation}
one has:
\begin{eqnarray} 
   \langle N-1,s | \hat{c}_m | N,0\rangle &=& 
   \sum_n \int d\mathbf{x} \,\phi^*_m(\mathbf{x}n) \,  
   f_s(\mathbf{x}n)
   \nonumber \\[5pt]
   &=& \langle \phi_m | f_s \rangle,
\end{eqnarray}
showing that a Dyson orbital $f_s(\mathbf{x})$ is null if and only if the amplitudes $\langle N-1,s | \hat{c}_m | N,0\rangle$ are zero for all $m$.
In making the above statement we have made use of Eq.~\eqref{eq:EOM_bath}, which connects the value of $f_s(\mathbf{x})$ in the physical subsystem to that of $f_s(\mathbf{x}n)$ for $n>0$ in the extended space. In fact, when the bath is coupled to the physical subsystem, $f_s(\mathbf{x})=0$ implies also $f_s(\mathbf{x}n)=0$.

For discussion purposes, suppose $\bar{O}$ is a (one-body) observable operator commuting with the Hamiltonian $\bar{H}$ in Eq.~\eqref{eq:full_hamiltonian}, that we can write as
\begin{equation}
   \bar{O} = \sum_m o_m \hat{c}^\dagger_m \hat{c}_m 
\end{equation}
on the $\{\phi_m(\mathbf{x}n)\}$ basis that diagonalizes $ \bar{O}$.
By evaluating $\langle N-1,s | \bar{O} \,\hat{c}_m | N,0\rangle$
and taking into account that $|N-1,s\rangle $ and $|N,0\rangle$ are simultaneous eigenvectors of $\bar{H}$ and $\bar{O}$ (with eigenvalues $O^{N-1}_s$ and $O^N_0$, respectively), one obtains:
\begin{equation}
   (O^{N-1}_s - O^N_0 -o_m) \, \langle N-1,s | \hat{c}_m | N,0\rangle = 0 .
\end{equation}
Similarly, considering a symmetry operator $\bar{S}=e^{i \bar{O}}$ (with any constant scalar factor adsorbed in $\bar{O}$) such that $[\bar{H},\hat{S}]=0$, by evaluating $\langle N-1,s | \bar{S}^\dagger \bar{S} \,\hat{c}_m | N,0\rangle$ one has:
\begin{equation}
   e^{i(O^{N-1}_s - O^N_0 -o_m)} \, \langle N-1,s | \hat{c}_m | N,0\rangle = 0 .
\end{equation}

The above expressions show that when $O^{N-1}_s$ and $O^N_0$ cannot be matched by any $o_m$ (when needed, modulo $2\pi$), all $m$ Dyson amplitudes need to be zero, thereby making $f_s(\mathbf{x})$ a null orbital. In other terms, null Dyson orbitals arise when $|N-1,s\rangle $ and $|N,0\rangle$ belong to two symmetry subspaces that cannot be connected by a single field operator. This is, e.g., the case of spin symmetry, where spin-0 states with $N$ particles, such as the ground state $|N,0\rangle$, can only form regular Dyson orbitals with spin-1/2 states having $N-1$ particles.

This discussion highlights that states $|N,m\rangle$ can be partitioned into those coupling to $|N-1,s\rangle$ corresponding to a regular Dyson orbital (by definition this is the case for the ground state $|N,0\rangle$), and those that do not couple to regular $|N-1,s\rangle$, due to symmetry reasons. This property is going to be used in the following to show that null Dyson orbitals do not contribute to $\delta v_{\text{ext}}(\mathbf{r},\epsilon_s)$ for $s$ any regular orbital.

%
\subsection{contributions to $\delta v_{\text{ext}}$}
By considering the EOM of Dyson orbitals, Eq.~\eqref{eq:dyson_r0_embed}, 
for the case of a null orbital (labeled by $t$ in the following), the following expression holds:
\begin{eqnarray}
   \label{eq:null_equation}
   \sum^R_{s'} K_{ts'}(\mathbf{r}) f_{s'}(\mathbf{r}) = 0 \qquad \qquad t\in Z .
\end{eqnarray}
This is going to be used in the following to rewrite $\delta v_{\text{ext}}(\mathbf{r},\epsilon_s)$ in Eq.~\eqref{eq:dvext_eom}.
%
%
In particular, 
by defining
\begin{eqnarray}
\delta v_{\text{ext}}(\mathbf{r},\epsilon_s) f_s(\mathbf{r}) &=& \phantom{-} P^{(1)}_s(\mathbf{r}) + P^{(2)}_s(\mathbf{r}),
\\[5pt]
\nonumber
P^{(1)}_s(\mathbf{r}) &=& 
     -\sum^R_{s'} \delta K_{ss'}(\mathbf{r}) \, f_{s'}(\mathbf{r}), 
\\
\nonumber
P^{(2)}_s(\mathbf{r}) &=& 
     -\sum^Z_{t}  K_{st}(\mathbf{r}) \, \delta f_{t}(\mathbf{r}),
\end{eqnarray}
one has:
\begin{widetext}

\begin{eqnarray}
P^{(1)}_s(\mathbf{r})
   &=& -\sum^R_{s'} \int d\mathbf{r}' v_c(\mathbf{r},\mathbf{r}')  \, 
   \Big[ \langle \delta N-1,s | \, \hat{n}(\mathbf{r}') \, | N-1,s'\rangle + \langle N-1,s | \,\hat{n}(\mathbf{r}') \,| \delta  N-1,s'\rangle
   \Big] \, f_{s'}(\mathbf{r}),
\\[5pt]   
&=& \phantom{-} \sum_{s'}^R \int d\mathbf{r}' v_c(\mathbf{r},\mathbf{r}') \sum^{\text{all}}_{p} 
\Bigg[ 
\frac{\langle N-1,s | \delta \bar{V}_{\text{ext}}| N-1,p\rangle}{ E^{N-1}_p -E^{N-1}_s} \, \langle N-1,p | \hat{n}(\mathbf{r}') | N-1,s'\rangle\, (1-\delta_{ps})
\label{eq:P1_line1}
\\ \nonumber
& & \qquad \qquad \qquad \qquad \qquad 
+(1-\delta_{ps'}) \, \langle N-1,s | \hat{n}(\mathbf{r}') | N-1,p\rangle\ \frac{\langle N-1,p | \delta \bar{V}_{\text{ext}}| N-1,s'\rangle}{ E^{N-1}_p -E^{N-1}_{s'}} \Bigg] f_{s'}(\mathbf{r}),
\\[5pt] 
\label{eq:P1_line2}
&=& 
 \sum_{s'}^R \sum^{R}_{p\neq s} \int d\mathbf{r}' v_c(\mathbf{r},\mathbf{r}')  
\frac{\langle N-1,s | \delta \bar{V}_{\text{ext}}| N-1,p\rangle}{ E^{N-1}_p -E^{N-1}_s} \, \langle N-1,p | \hat{n}(\mathbf{r}') | N-1,s'\rangle\, f_{s'}(\mathbf{r}) 
\\ \nonumber
& & \quad \!\! +
\sum^{\text{all}}_p \int d\mathbf{r}' v_c(\mathbf{r},\mathbf{r}')  \,
\langle N-1,s | \hat{n}(\mathbf{r}') | N-1,p\rangle\ 
\langle \delta N-1,p | \hat{\psi}(\mathbf{r}) | N,0\rangle, 
\\[7pt]
   \label{eq:P2_def}
   P^{(2)}_s(\mathbf{r}) 
     &=& -\sum_{t}^{Z} \int d\mathbf{r}' v_c(\mathbf{r},\mathbf{r}')  \, \langle N-1,s | \hat{n}(\mathbf{r}') | N-1,t\rangle  \Big[ \langle \delta N-1,t | \hat{\psi}(\mathbf{r}) | N,0\rangle + \langle N-1,t | \hat{\psi}(\mathbf{r}) | \delta N,0\rangle \Big].
\end{eqnarray}
\end{widetext}
In passing from Eq.~\eqref{eq:P1_line1} to \eqref{eq:P1_line2} we have first made use of Eq.~\eqref{eq:null_equation} to limit the sum over $p$ to regular Dyson orbitals, and then adsorbed the sum over $s'$ in the second term to obtain $\langle \delta N-1,p |$.

At this point one can already appreciate that in the first term of Eq.~\eqref{eq:P1_line2}, all excited states at $N-1$ particles are limited to the regular set $R$ ($f_s$ is also regular by construction), so that the matrix elements $\langle N-1,s | \delta \bar{V}_{\text{ext}}| N-1,p\rangle$ are zero by virtue of Eq.~\eqref{eq:key_zero_mat_element2}.
The whole first term can then be neglected. In the following we pursue this strategy further.
For instance, the term $P^{(2)}$ in Eq.~\eqref{eq:P2_def}
cancels the $Z$ contribution to the second line of Eq.~\eqref{eq:P1_line2}, thereby also limiting that sum to $R$.
%
%
Expanding also $|\delta N,0\rangle$ using perturbation theory, 
the overall variation of the embedding potential in Eq.~\eqref{eq:dvext_eom} can be written as:
\begin{widetext}
\begin{eqnarray}
\label{eq:dvexp_final}
\delta v_\text{ext}(\mathbf{r},\epsilon_s) f_s(\mathbf{r}) 
&=& 
 \sum_{s'}^R \sum^{R}_{p\neq s} \int d\mathbf{r}' v_c(\mathbf{r},\mathbf{r}')  
\frac{\langle N-1,s | \delta \bar{V}_{\text{ext}}| N-1,p\rangle}{ E^{N-1}_p -E^{N-1}_s} \, \langle N-1,p | \hat{n}(\mathbf{r}') | N-1,s'\rangle\, f_{s'}(\mathbf{r}) 
\\ \nonumber
&+&
\sum_{s'}^R \sum^{R}_{p\neq s'} \int d\mathbf{r}' v_c(\mathbf{r},\mathbf{r}')  
\langle N-1,s | \hat{n}(\mathbf{r}') | N-1,p\rangle\,
\frac{\langle N-1,p | \delta \bar{V}_{\text{ext}}| N-1,s'\rangle}{ E^{N-1}_p -E^{N-1}_{s'}} \, f_{s'}(\mathbf{r}) 
\\ \nonumber
&+& 
 \sum_{t}^Z \sum_{m>0} \int d\mathbf{r}' v_c(\mathbf{r},\mathbf{r}')  
\langle N-1,s | \hat{n}(\mathbf{r}') | N-1,t\rangle\,
\langle N-1,t | \hat{\psi}(\mathbf{r}) | N,m\rangle\,
\frac{\langle N,m | \delta \bar{V}_{\text{ext}}| N,0\rangle}{ E^{N}_m -E^{N}_0} .
\end{eqnarray}
\end{widetext}
As discussed above, the first two terms are zero because involve sums over $N-1$ excited states that are both regular, so that Eq.~\eqref{eq:key_zero_mat_element2} can be invoked. 
It remains to discuss the term in the third line, involving the matrix elements $\langle N,m | \delta \bar{V}_{\text{ext}}| N,0\rangle$.

Making reference to the symmetry discussion made at the beginning of this Section (including the case of spin-1/2 interacting Fermions),  
null Dyson orbitals originate from a system symmetry such that $|N-1,t\rangle$ that does not couple to the ground state $|N,0\rangle$.
Under this conditions we see that $|N,m\rangle$ should have a spin-symmetry different from $|N,0\rangle$ in order to give a non-zero matrix element $\langle N-1,t|\hat{\psi}(\mathbf{r}) | N,m\rangle$. According to Eq.~\eqref{eq:Nm_dvext_N0}, this is nevertheless incompatible with having non-zero matrix elements $\langle N,m | \delta \bar{V}_{\text{ext}}| N,0\rangle$. In fact, the sum over $s$ in Eq.~\eqref{eq:Nm_dvext_N0} can be limited to regular Dyson orbitals and, in order to have non negligible matrix elements, one needs $|N,m\rangle$ to have the same symmetry of the ground state and couple to regular $|N-1,s\rangle$ wavefunctions. As a result, either $\langle N,m | \delta \bar{V}_{\text{ext}}| N,0\rangle=0$ or $\langle N-1,t | \hat{\psi}(\mathbf{r}) | N,m\rangle = 0$.


\section{Dynamical Sham-Schl\"uter equations}
\label{sec:app_dsse}
%
In this Section we discuss how to compute $v^{\text{SP}}(\mathbf{r},\omega)$ by means of dynamical Sham-Schl\"uter equations~\cite{Gatti2007PhysRevLett,Ferretti2014PhysRevB, Vanzini2020FaradayDiscuss}, which we here adapt and specialize to the present formulation of the problem.

As a preliminary step, we discuss how to extract the spectral density from a given Green's function with discrete poles. In particular, taking $\Gamma_s$ to be a contour in the complex frequency plane small enough to encircle only the pole $\epsilon_s$, according to Eqs.~\eqref{eq:lehmann} and ~\eqref{eq:GSP_def} one has
\begin{equation}
\label{eq:get_spectral_rho}
  \oint_{\Gamma_s} \frac{dz}{2 \pi i} \,G^{(\text{SP})}(\mathbf{r},\mathbf{r},z) = \rho_s(\mathbf{r}),
  \qquad \qquad \forall s \in \text{occ},
\end{equation}
which is valid for both $G$ and $G^{\text{SP}}$.
One also has:
\begin{eqnarray}
    & &\int d\mathbf{r} \oint_{\Gamma_s} \frac{dz}{2 \pi i} \,G^{(\text{SP})}(\mathbf{r},\mathbf{r},z) = Z_s,
    \nonumber \\
    & &\int d\mathbf{r} \oint_{\Gamma_s} \frac{dz}{2 \pi i} \,z G^{(\text{SP})}(\mathbf{r},\mathbf{r},z) = \epsilon_s Z_s,   
    \qquad \qquad
\end{eqnarray}
which are nevertheless both contained in Eq.~\eqref{eq:get_spectral_rho} since $Z_s$ is the integral of $\rho_s(\mathbf{r})$ and $\epsilon_s$ is implied by the definition of $\Gamma_s$.
While Eqs.~\eqref{eq:get_spectral_rho}, when applied to $G^{\text{SP}}$, could be used to determine $v^{\text{SP}}$, the potential enters in the determination of $G^{\text{SP}}$ via the inversion operation of Eq.~\eqref{eq:GSP_def}. Instead, one can follow the usual approach leading to SSEs,  Ref.~\cite{Gatti2007PhysRevLett} in particular, and apply Eq.~\eqref{eq:get_spectral_rho} to the Dyson equation connecting $G$ and $G^\text{SP}$, 
\begin{equation}
   G = G^{\text{SP}} + G^{\text{SP}} \left[  \Sigma_{\text{Hxc}} -v^{\text{SP}}_{\text{Hxc}}\right] G, 
\end{equation}
where we have written $v^{\text{SP}}(\mathbf{r},z) = v_{\text{ext}}(\mathbf{r},z) + v^{\text{SP}}_{\text{Hxc}}(\mathbf{r},z)$ according to Eq.~\eqref{eq:vSP_decomposition}.
By doing so one obtains:
\begin{eqnarray}
   \oint_{\Gamma_s} \frac{dz}{2 \pi i} \int \!\!d\mathbf{r}_1\,G^{\text{SP}}(\mathbf{r},\mathbf{r}_1,z)  G(\mathbf{r}_1,\mathbf{r},z) \, v^{\text{SP}}_{\text{Hxc}}(\mathbf{r}_1,z) =
   \nonumber
   \\
    =\oint_{\Gamma_s} \frac{dz}{2 \pi i} 
    \langle \mathbf{r} | G^{\text{SP}}(z) \Sigma_{\text{Hxc}}(z) G(z) | \mathbf{r} \rangle 
    \quad  \forall s \in \text{occ}. \quad
\end{eqnarray}
By making use of Eqs.~\eqref{eq:GSP_def2} and \eqref{eq:lehmann} it is possible to evaluate the contour integrals and one can eventually obtain a set of coupled equations in the unknowns $v^{\text{SP}}_{\text{Hxc}}(\mathbf{r},\epsilon_s)$. Explicitly, one has:
\begin{equation}
     \label{eq:dsse_linsys}
     \int d\mathbf{r}_1, K_s(\mathbf{r},\mathbf{r}_1,[G^\text{SP}]) \, v^{\text{SP}}_{\text{Hxc}}(\mathbf{r}_1,\epsilon_s) = B_s(\mathbf{r},[G^\text{SP}]),
\end{equation}
where we have defined:
\begin{widetext}
\begin{eqnarray}
   K_s(\mathbf{r},\mathbf{r}_1,[G^\text{SP}]) &=& \sum_{p\neq s}^{\text{all}} \frac{P_s^{\text{SP}}(\mathbf{r},\mathbf{r}_1) P_p(\mathbf{r},\mathbf{r}_1)}{\epsilon_s-\epsilon_p} + 
   \frac{P_p^{\text{SP}}(\mathbf{r},\mathbf{r}_1) P_s(\mathbf{r},\mathbf{r}_1)}{\epsilon_s-\epsilon_p},
   \\
   B_s(\mathbf{r},[G^\text{SP}]) &=& 
   \sum_{p\neq s}^{\text{all}} \int d\mathbf{r}_1 d\mathbf{r}_2 \, \left[
       \frac{P_s^{\text{SP}}(\mathbf{r},\mathbf{r}_1) P_p(\mathbf{r}_2,\mathbf{r})}{\epsilon_s-\epsilon_p} + 
   \frac{P_p^{\text{SP}}(\mathbf{r},\mathbf{r}_1) P_s(\mathbf{r}_2,\mathbf{r})}{\epsilon_s-\epsilon_p}
   \right] \Sigma_{\text{Hxc}}(\mathbf{r}_1,\mathbf{r}_2,\epsilon_s) 
   \\[5pt] \nonumber
   &+& 
   \int d\mathbf{r}_1 d\mathbf{r}_2 \, P_s^{\text{SP}}(\mathbf{r},\mathbf{r}_1) \left[ \dot{\Sigma}_{\text{Hxc}}(\mathbf{r}_1,\mathbf{r}_2,\epsilon_s) -\dot{v}^{\text{SP}}_{\text{Hxc}}(\mathbf{r}_1,\epsilon_s)\delta_{\mathbf{r}_1,\mathbf{r}_2} \right]  P_s(\mathbf{r}_2,\mathbf{r}),
\end{eqnarray}
\end{widetext}
and we have set $P^{(\text{SP})}_s(\mathbf{r},\mathbf{r}') = f^{(\text{SP})}_s(\mathbf{r})\,f^{(\text{SP})*}_s(\mathbf{r}')$.

Given the non-linearity of the equations, one can proceed iteratively by first making a trial guess about $v^{\text{SP}}(\mathbf{r},\omega)$, then computing $G^{\text{SP}}$ and $\dot{v}^{\text{SP}}$ and evaluating $K_s$ and $B_s$, and finally solving the linear systems in Eq.~\eqref{eq:dsse_linsys} for each $s$ corresponding to occupied Dyson orbitals. Once all $v^{\text{SP}}(\mathbf{r},\epsilon_s)$ are known, one can combine the information with the knowledge of $\dot{v}^\text{SP}_s$ (which can be obtained directly from $Z_s$), to interpolate a new unique $v^{\text{SP}}_{\text{Hxc}}(\mathbf{r},\omega)$ according to the discussion in App.~\ref{sec:math_uniqueness_rational}. The procedure is the repeated until convergence. The equations above will serve as a starting point for further analytical and numerical studies.

%
%
\renewcommand{\emph}{\textit}

%

\end{document}